\documentclass[aps,prl,twocolumn,superscriptaddress,footinbib,floatfix,showkeys,longbibliography]{revtex4-2}
\usepackage{graphicx}  % Include figure files
\usepackage{subfigure}
\usepackage{dcolumn} % Align table columns on decimal point
\usepackage{bm}% bold math
\usepackage{amssymb}   % for math
\usepackage{amsfonts}   % for math
\usepackage{comment}
\usepackage[colorlinks,linkcolor=blue,anchorcolor=blue,citecolor=blue,urlcolor=blue]{hyperref}
\usepackage{xcolor} 
\usepackage{soul}
\usepackage{tabularx,booktabs,ragged2e,multirow}
\newcolumntype{C}{>{\centering\arraybackslash}X}
\usepackage{multirow}
\usepackage{amsmath}
\usepackage{mathrsfs}
\usepackage{mathcomp}
\usepackage{textcomp}
\usepackage{dsfont}
\usepackage{esint}
\usepackage{braket}
\usepackage{textgreek}
\usepackage{lipsum}
\usepackage{marvosym} 
\usepackage{centernot}
\usepackage{cancel}
\usepackage{dashrule}
\usepackage{tikz}

\usepackage{stackengine}
\stackMath

\DeclareMathOperator{\Det}{Det}

\DeclareMathOperator{\sgn}{sgn}

\begin{document}
%\preprint{APS/123-QED}

%\title{Low-energy signature and magnetic suppression of NH skin effects}
\title{Surface spectroscopy and surface-bulk hybridization of Weyl semimetals}

\author{Xiao-Xiao Zhang}
%\email{xxzhang@hust.edu.cn}
\affiliation{Wuhan National High Magnetic Field Center and School of Physics, Huazhong University of Science and Technology, Wuhan 430074, China}
%\affiliation{Institute for Quantum Science and Engineering, Huazhong University of Science and Technology, Wuhan 430074, China}
%\affiliation{Wuhan Institute of Quantum Technology, Wuhan 430206, China}
\affiliation{RIKEN Center for Emergent Matter Science (CEMS), Wako, Saitama 351-0198, Japan}

\author{Naoto Nagaosa}
%\email{nagaosa@ap.t.u-tokyo.ac.jp}
%\affiliation{Department of Applied Physics, University of Tokyo, Tokyo 113-8656, Japan}
\affiliation{RIKEN Center for Emergent Matter Science (CEMS), Wako, Saitama 351-0198, Japan}

%\date{\today}

\newcommand{\ba}{{\bm a}}
\newcommand{\bd}{{\bm d}}
\newcommand{\bb}{{\bm b}}
\newcommand{\bk}{{\bm k}}
\newcommand{\bmm}{{\bm m}}
\newcommand{\bn}{{\bm n}}
\newcommand{\br}{{\bm r}}
\newcommand{\bq}{{\bm q}}
\newcommand{\bp}{{\bm p}}
\newcommand{\bu}{{\bm u}}
\newcommand{\bv}{{\bm v}}
\newcommand{\bA}{{\bm A}}
\newcommand{\bB}{{\bm B}}
\newcommand{\bD}{{\bm D}}
\newcommand{\bE}{{\bm E}}
\newcommand{\bH}{{\bm H}}
\newcommand{\bJ}{{\bm J}}
\newcommand{\bK}{{\bm K}}
\newcommand{\bL}{{\bm L}}
\newcommand{\bM}{{\bm M}}
\newcommand{\bP}{{\bm P}}
\newcommand{\bR}{{\bm R}}
\newcommand{\bS}{{\bm S}}
\newcommand{\bX}{{\bm X}}
\newcommand{\brho}{{\bm \rho}}
\newcommand{\cA}{{\mathcal A}}
\newcommand{\cB}{{\mathcal B}}
\newcommand{\cC}{{\mathcal C}}
\newcommand\cD{\mathcal{D}}
\newcommand{\cE}{{\mathcal E}}
\newcommand{\cG}{{\mathcal G}}
\newcommand{\cH}{{\mathcal H}}
\newcommand{\cK}{{\mathcal K}}
\newcommand{\cM}{{\mathcal M}}
\newcommand{\cP}{{\mathcal P}}
\newcommand{\cT}{{\mathcal T}}
\newcommand{\cV}{{\mathcal V}}
\newcommand{\balpha}{{\bm \alpha}}
\newcommand{\bbeta}{{\bm \beta}}
\newcommand{\bdelta}{{\bm \delta}}
\newcommand{\bgamma}{{\bm \gamma}}
\newcommand{\bGamma}{{\bm \Gamma}}
\newcommand{\bzero}{{\bm 0}}
\newcommand{\bOmega}{{\bm \Omega}}
\newcommand{\bsigma}{{\bm \sigma}}
\newcommand{\bUpsilon}{{\bm \Upsilon}}
\newcommand{\bcA}{{\bm {\mathcal A}}}
\newcommand{\bcB}{{\bm {\mathcal B}}}
\newcommand{\bcD}{{\bm {\mathcal D}}}
\newcommand{\Domega}{ {\Delta\omega}}
\newcommand{\Dbk}{ {\Delta\bk}}
\newcommand\dd{\mathrm{d}}
\newcommand\ii{\mathrm{i}}
\newcommand\ee{\mathrm{e}}
\newcommand\bark{\bar{k}}
\newcommand\bbark{\bar{\bm k}}
\newcommand\zz{\mathsf{z}}
\newcommand\colonprod{\!:\!}

\makeatletter
\let\newtitle\@title
\let\newauthor\@author
\def\ExtendSymbol#1#2#3#4#5{\ext@arrow 0099{\arrowfill@#1#2#3}{#4}{#5}}
\newcommand\LongEqual[2][]{\ExtendSymbol{=}{=}{=}{#1}{#2}}
\newcommand\LongArrow[2][]{\ExtendSymbol{-}{-}{\rightarrow}{#1}{#2}}
\newcommand{\cev}[1]{\reflectbox{\ensuremath{\vec{\reflectbox{\ensuremath{#1}}}}}}
\newcommand{\red}[1]{\textcolor{red}{#1}} %for displaying red texts
\newcommand{\blue}[1]{{\leavevmode\color{blue}#1}} %for displaying blue texts
\newcommand{\green}[1]{\textcolor{orange}{#1}} %for displaying blue texts
\newcommand{\mytitle}[1]{\textcolor{orange}{\textit{#1}}}
\newcommand{\mycomment}[1]{} %for commenting out
\newcommand{\note}[1]{ \textbf{\color{blue}#1}}
\newcommand{\warn}[1]{ \textbf{\color{red}#1}}

\makeatother

\begin{abstract}
Weyl semimetal showing open-arc surface states is a prominent example of topological quantum matter in three dimensions. With the bulk-boundary correspondence present, nontrivial surface-bulk hybridization is inevitable but less understood.
Spectroscopies have been often limited to verifying the existence of surface Fermi arcs, whereas its spectral shape related to the hybridization profile in energy-momentum space is not well studied. 
We present an exactly solvable formalism at the surface for a wide range of prototypical Weyl semimetals. The resonant surface state and the bulk influence coexist as a surface-bulk hybrid and are treated in a unified manner. Directly accessible to angle-resolved photoemission spectroscopy, we analytically reveal universal information about the system obtained from the spectroscopy of resonant topological states. We systematically find inhomogeneous and anisotropic singular responses around the surface-bulk merging borderline crossing Weyl points, highlighting its critical role in the Weyl topology. The response in scanning tunneling spectroscopy is also discussed. 
The results will provide much-needed insight into the surface-bulk-coupled physical properties and guide in-depth spectroscopic investigation of the nontrivial hybrid in many topological semimetal materials.
\end{abstract}
%\pacs{71.10.Pm, 71.27.+a, 72.15.Nj, 72.15.Rn}
%\keywords{quantum Hall effect, Weyl semimetal, magnetotransport, anisotropy, three dimensions}

\maketitle
% \tableofcontents
% \newpage
% \clearpage

%\clearpage
\let\oldaddcontentsline\addcontentsline% Store \addcontentsline
\renewcommand{\addcontentsline}[3]{}% Make \addcontentsline a no-op

\subsection*{Significance statement}
Weyl semimetal is an epitome of frontier topological quantum materials. Its surface property accessible to photoemission measurements is important and contains rich information about the system. Lacking analysis of the inherent three-dimensional surface-bulk hybrid nature, investigations are often limited to verifying the existence of surface states and miss the hybridization aspect of general bulk-boundary correspondence. Based on an exactly solvable formalism consistently treating the surface and bulk states, we reveal the missing mathematical structure of the surface-bulk hybrid and the hidden information about the topological surface state; singular spectroscopic responses of the surface-bulk merging are highlighted. This underscores the importance of appreciating the intricate surface-bulk hybridization in quantum materials, paving the way for exploring such unique features.

\section*{Introduction}
Topological quantum materials, characteristically accompanied by nontrivial topological states at the boundary, are an important frontier of condensed matter physics.
Intriguing physical properties are often closely related to the principle of bulk-boundary correspondence and the boundary states. 
An epitome in three dimensions (3D) is the Weyl semimetal (WSM), which is significantly beyond an extension of the two-dimensional (2D) Dirac physics\cite{Volovik1987,Weyl2007,Weyl2011,AHE2,WeylwithP1,predict1,predict2,TaAs2,TaAs1,TaAs4,TaAs3,WeylDiracReview,Lv2021}. 
It shows Weyl points (WPs) with linear band crossings as momentum-space Dirac monopoles and %chiral magnetic effect and negative magnetoresistance due to 
the associated chiral anomaly effects\cite{Nielson-Ninomiya2,CME1,Burkov2014,seeCMEDirac1,seeCMEDirac2,seeCMEWeyl2,Cheng2021,Ong2021}. 
At its boundary, unusual open-arc surface states appear in the projected surface Brillouin zone (BZ)\cite{Burkov2016,ReviewYan2017,Burkov2018,WeylDiracReview,Nagaosa_2020,Lv2021}. However, such Fermi arc surface states cannot be separated from the bulk at will, e.g., in transport measurements, because they interplay to form an organic whole and hence contribute together. For instance, unique quantum oscillations and quantum Hall effects originate from the Weyl orbit combining arc and bulk states\cite{Potter2014,Zhang2016b,Borchmann2017,Wang2017a,Zhang2018,Nishihaya2021,Zhang2021a}. An important question arises: what are the mathematical structure and physical effects of the inherent 3D surface-bulk hybrid in WSMs?

How the boundary and bulk are nontrivially connected and hybridized is a less clear but indispensable aspect of the general principle of bulk-boundary correspondence. WSMs provide an ideal platform for such investigation because their 2D surfaces are open to many experimental probes.
However, spectroscopy has been often limited to merely verifying the existence of surface states and overlooked the bulk-boundary hybrid and its nontrivial energy-momentum profile.
Such information actually lies in the neighborhood of the surface and hides rich information about the whole system, crucial for studying further optical responses, transport and magnetic properties\cite{Inoue2016,Gorbar2016,Duan2018,Cheng2020,Nagaosa_2020}. 
It should be naturally detectable by contemporary spectroscopies, e.g., the angle-resolved photoemission spectroscopy (ARPES) that resolves the important momentum dependence and the scanning tunneling spectroscopy (STS) that encodes the density of states in the tunneling I-V characteristics\cite{Lv2019,Sobota2021,Binnig1987,Bonnell2000}. Photoemission is presumably more tunable between surface or bulk sensitive via the penetration depth of variable photon energy, which thus can accommodate sensitively the effect of surface-bulk hybrid\cite{Petersen2000,Deng2016}.
Further equipped with spin resolution, these techniques can extract information inherent to the spin channel and magnetism as demonstrated experimentally\cite{Cacho2015,Jozwiak2016,Lv2019,Sobota2021,Bode2003,Wiesendanger2009}.
%Such photoemission typically averages over several atomic layers and thus can accommodate more sensitively the effect of surface-bulk hybrid.

%This is mainly caused by the lack of concrete analysis of the 3D surface-bulk coupled system and its full spectroscopic implications. 
To this end, we present a general formalism for a wide range of typical WSMs (in terms of symmetry, number of WPs, and type of Weyl fermion) with an exact solution near the surface. The consistently unified surface state and bulk influence substantially broaden the physical reach.
Targeting surface ARPES measurements, we find that the surface resonance signal contains much more information about the whole system than usually deemed, hidden but comprehensively encoded in its shape and intensity profile. A byproduct is finding unequal numbers of Fermi arcs and WP pairs, contrary to the common belief.
Revealing the remarkable structure of surface-bulk hybrid, highly anisotropic singular responses near the loop-like surface-bulk merging borderline in the energy-momentum space are found accompanied by anomalous inverse square root scaling behavior. They provide a concrete prediction and picture of the borderline crucially mediating the surface-bulk hybridization, which serves as a critical transition boundary between surface and bulk effects, together with its own special edge at WPs.
Due to the spin-polarized surface state, spin-resolved STS measurement can also achieve a separation of the bulk and surface tunneling contributions.

\section*{Results}

\subsection*{Model and effective surface Green's function}
\begin{figure}[!hbt]
\centering
\includegraphics[width=8.7cm]{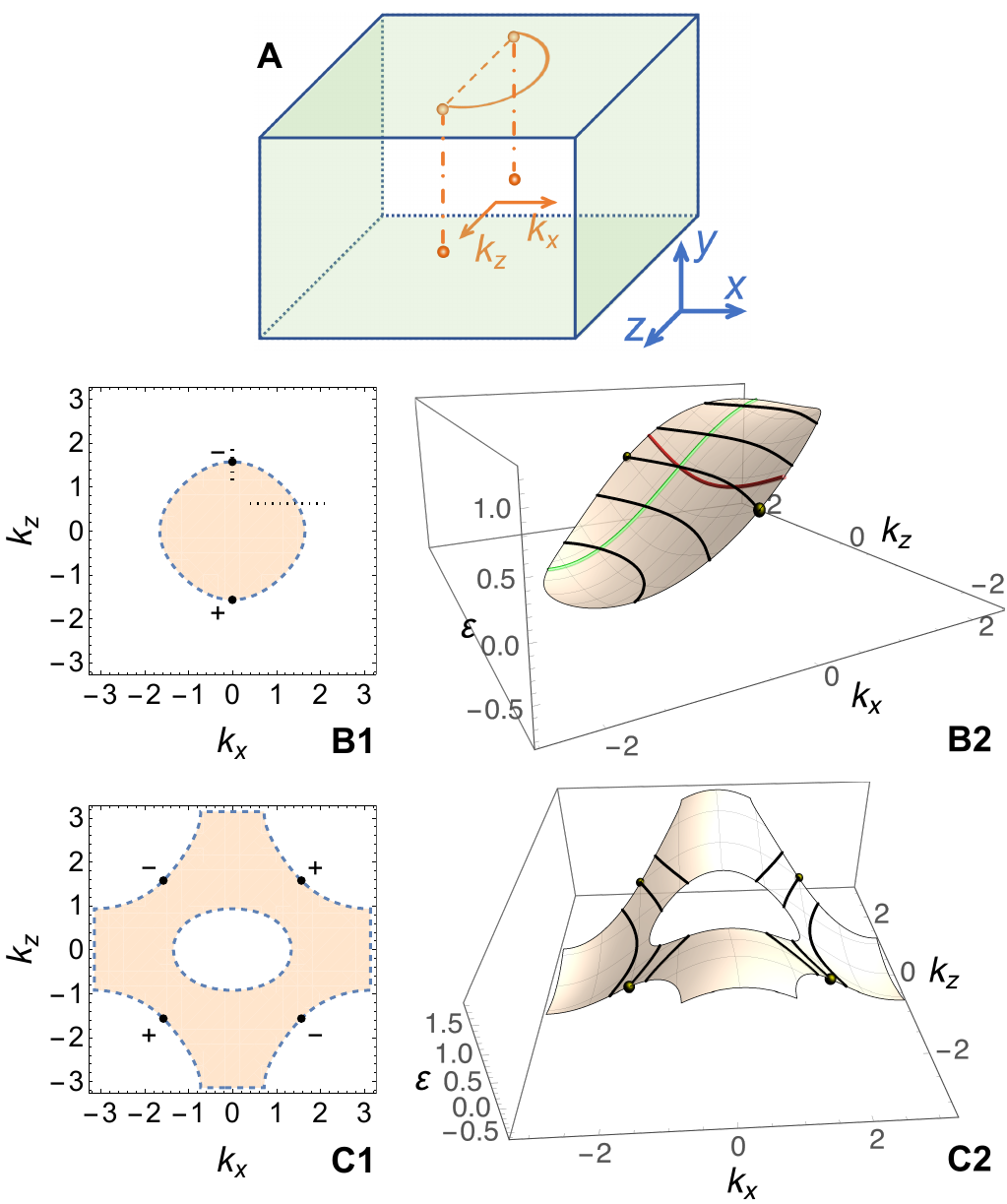}
\caption{Topological surface state on the top surface of a WSM. (A) Schematic of the representative minimal model system (systems with more WPs can also be treated). Orange coordinates and shapes additionally indicate the momentum space. One WP pair along $k_z$-axis is projected onto the 2D top surface Brillouin zone (BZ), where a solid Fermi arc state appears. (B1) The orange momentum constraint region of the surface state in the surface BZ. WP charges $\pm$ are noted. Horizontal and vertical dotted lines correspond to the selected paths in Fig.~\ref{Fig:Aomegakxkz3D_main}. (B2) Surface state dispersion $\varepsilon(\bk)$ restricted in the constraint region. Projected WP pair is indicated by black dots. Green and red lines are momentum cuts of the surface state at fixed $k_z$ or $k_x$, respectively corresponding to Fig.~\ref{Fig:Aomegakxkz_main}(A,B).
Black lines are isoenergy lines of the surface state, i.e., Fermi arcs, corresponding to arc state signals at various Fermi energies in Fig.~\ref{Fig:AkxkzArc_main} from low to high energies. Full parameters are given in Fig.~\ref{Fig:Aomegakxkz_main}.
Panels (C1,C2) present a different noncentrosymmetric model with two pairs of WPs, which bears a multiply connected surface state constraint region. Within a certain range of $E_F$, e.g., at the WP energy, there appear two Fermi arcs for one WP pair due to the inner hole structure. 
}
\label{fig:SSreg-band_main}
\end{figure}

We consider a two-band Hamiltonian that breaks both the time-reversal and inversion symmetries 
\begin{equation}\label{eq:H_main}
\begin{split}
    &\cH(\bk)= [D_x (1-\cos{k_x})+D_z (1-\cos{k_z})]\sigma_0\\
    &-t \sum_{i=x,y,z}(\cos{k_i}-\cos{\bark_i})\sigma_1\\
    &+ (t\sin{k_y}-t_2\cos{k_x})\sigma_2+t_1\sin{k_x}\sigma_3.
\end{split}
\end{equation}
With appropriate parameter ranges, it gives a minimal pair of WPs $\bbark^\pm=(0,\sin^{-1}\frac{t_2}{t},\pm k_w)$ aligned along the $k_z$-direction as shown in Fig.~\ref{fig:SSreg-band_main}(A) and at the energy $\varepsilon_0=D_z(1-\cos{k_w})$. 
Finite $D_x,D_z$ break the particle-hole symmetry and add extra spin-independent dispersion in the $xz$-plane surface of our main interest; especially $D_z$ accounts for the realistic Fermi arcs with finite curvature, which is common in real materials and crucial, e.g., to many phenomena involving Weyl orbits.
We show how this Hamiltonian is formulated and related to more general cases in Supplemental Information (SI) Sec.~\ref{SM:Hamiltonian}. 
Also discussed later, most other types of WSMs with different symmetries and more WPs can be treated.

The surface effective Green's function can be analytically obtained by considering a sample semi-infinite in the $-\hat{y}$-direction. The Hamiltonian of the top surface, if decoupled from the bulk, is given by (summation over $\alpha=0,\cdots,3$ is henceforth understood) $h=d_\alpha(\bk)\sigma_\alpha$, where $d_0=D_x (1-\cos{k_x})+D_z (1-\cos{k_z}),d_1=-t (\sum_{i=x,z}\cos{k_i}-\sum_{i=x,y,z}\cos{\bark_i}),d_2=-t_2\cos{k_x},d_3=t_1\sin{k_{x}}$. We use $\bk$ to denote $k_x,k_z$-dependence in the surface BZ for simplicity, which is our main focus.
The whole system is then cast in a recursive form of $h$ coupled to neighboring layers.
With the shorthand notation $\bar d_0=z-d_0$ for generic complex frequency $z$ and $d_{12}=\sqrt{d_1^2+d_2^2}$, we obtain \textit{two} exact solutions labelled by $r=\pm1$ for the surface Green's function $g(z,\bk)$ (\textit{Methods})
\begin{equation}\label{eq:ginv&g}
\begin{split}
    &g=b_0\sigma_0+\bb\cdot\bsigma,\quad b_0-b_3=-\frac{(\bar d_0-d_3)(K+2t^2)}{2d_{12}^2t^2},\\
    &b_{1,2}=-\frac{d_{1,2}(K+2t^2)}{2d_{12}^2t^2},\quad b_0+b_3=-\frac{K}{2(\bar d_0-d_3)t^2},
\end{split}
\end{equation}
where 
\begin{equation}\label{eq:Ks}
    K(r)=B+r\sqrt{C}
\end{equation}
depends on the choice of $r$ with 
\begin{equation}\label{eq:C}
    C=B^2+4t^2F
\end{equation}
and $B=d^2-\bar d_0^2- t^2,F=d_3^2-\bar d_0^2$.
Momentum and energy dependence is henceforth suppressed for brevity.
This solution is found by effectively integrating out the bulk degrees of freedom and the surface effective Green's function $g$ will encode the full information of the original WSM system. A related approach via integrating the bulk has been applied to mesoscopic transport calculation and quantum oscillation under magnetic field\cite{disorder:RecursiveGreen,Borchmann2017}.

\subsection*{Spectral function}\label{sec:spec}
\begin{figure*}[!htbp]
\centering
\includegraphics[width=17.8cm]{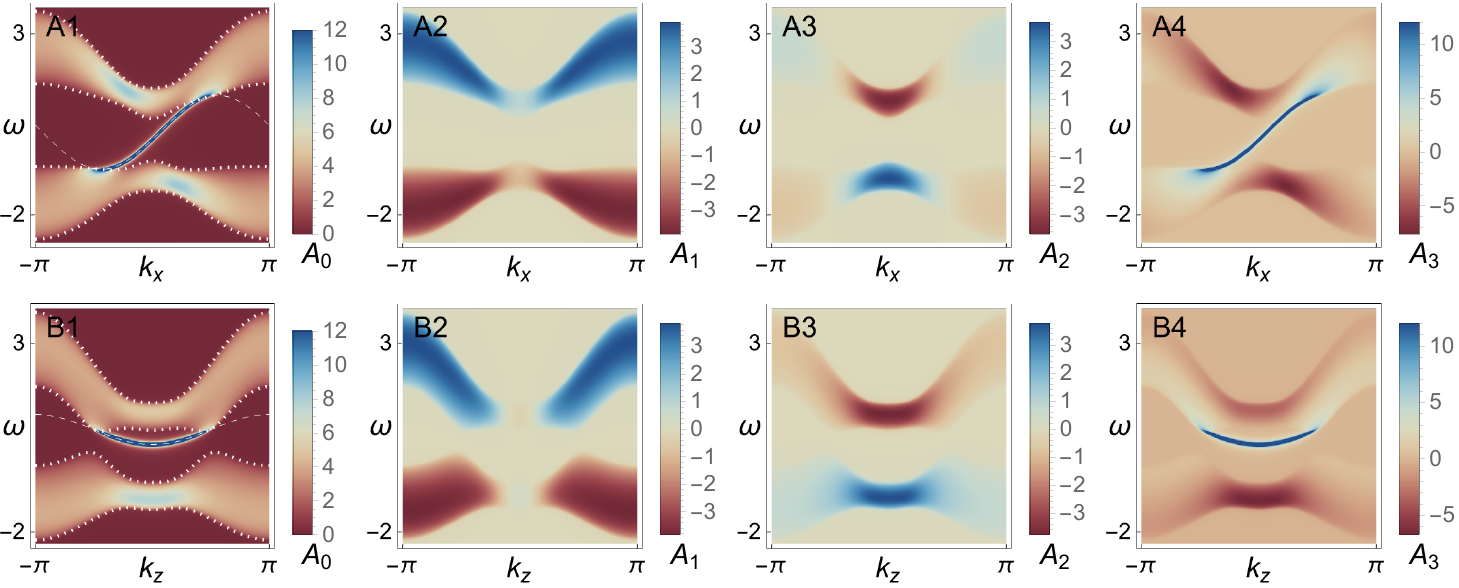}
\caption{Spin-resolved ARPES signals of the surface of a WSM in (A) the $(\omega,k_x)$-plane at $k_z=\pi/5$ and (B) the $(\omega,k_z)$-plane at $k_x=\pi/10$. Spectral function $A_{0,1,2,3}$ successively in the charge and three spin channels for the WSM with parameters $k_w=\pi/2,t=t_1=1,t_2=0.4,D_x=0.2,D_z=0.4,\eta=0.02$. 
A deep blue surface state (resonance intensity clipped to enhance the overall visibility) is seen only and identically in the $A_{0,3}$ channels, signifying the charge-spin locking between them. 
Such a surface state in panels (A1,A4) and (B1,B4) respectively corresponds to the green and red lines in Fig.~\ref{fig:SSreg-band_main}(B2).
Exemplified in the $A_0$ channel, it tangentially merges into the projected bulk states bounded by the dotted lines; for clarity, the full surface state dispersion beyond the momentum constraint region is indicated by dashed lines.
}\label{Fig:Aomegakxkz_main}
\end{figure*}

The information most relevant to spectroscopy can be extracted from the spectral function of Green's function $g(z,\bk)$ in \eqref{eq:ginv&g}. It is given by %(summation over $\alpha=0,\cdots,3$ is henceforth understood)
\begin{equation}\label{eq:cA}
    A(\omega,\bk)=\ii(g^\mathrm{r}-g^{\mathrm{r}\dag})=A_\alpha\sigma_\alpha
\end{equation}
where the retarded Green's function $g^\mathrm{r}=g(z\rightarrow\omega+\ii 0^+)$, leading to the charge ($\alpha=0$) and three spin ($\alpha=1,2,3$) channels
\begin{equation}\label{eq:A_alpha}
    A_\alpha(\omega,\bk)=-2\Im \,\mathrm{tr}[g^\mathrm{r}\sigma_\alpha]=-4\lim_{\eta\rightarrow0^+}\Im \, b_\alpha(z\rightarrow\omega+\ii \eta).
\end{equation}
In real systems, $\eta$ may signify the phenomenological strength of possible relaxation mechanisms, e.g., disorder, phonon scattering, electron correlation, etc.
ARPES without spin resolution measures $A_0$ only while spin-resolved ARPES can extract each of $A_\alpha$ with spin-polarized photoelectron detectors.
To understand its physical meaning, it is crucial to know how $r=\pm1$ in \eqref{eq:Ks} should be chosen, which turns out to be a nontrivial issue because $C$ of \eqref{eq:C} is \textit{not} necessarily always positive.
It is, however, remarkable to note that the above definition of spectral function holds in general \textit{regardless} of the sign of $C$ under the square root because Green's functions can generally be complex.

We plot the above spectral function in Fig.~\ref{Fig:Aomegakxkz_main}. Exemplified in Fig.~\ref{Fig:Aomegakxkz_main}(A1,B1), both the topological surface state and the influence from the bulk states are unified in $g$ and hence in $A$; the surface state eventually ends and merges into the bulk at a certain borderline in the energy-momentum space.
To facilitate and guide the discussion, we point out three crucial aspects of the key quantity $C$:
\begin{itemize}
    \item The sign of $C$ determines in the energy-momentum space the boundary between bulk contribution ($C<0$) and the region admissible to surface state ($C>0$);
    \item The borderline of surface state entails $C=0$, where the surface state merges into the bulk states;
    \item The choice of $r=\pm1$ has to be determined physically according to the positivity of the charge channel spectral function $A_0$, because it directly corresponds to the density of states.
\end{itemize}
Henceforth, when the sign of a certain quantity is concerned, complex frequency $z$ substituted by real $\omega$ is understood.

\subsection*{Surface resonance spectroscopy}
We first find the states on-shell at the surface, i.e., the topological surface state. As long as $C>0$ %and hence $K\in\mathds{R}$ 
for some energy-momentum region, 
the analytic continuation in \eqref{eq:A_alpha} matters only in the denominator and gives (\textit{Methods})
\begin{equation}\label{eq:specA}
    A_{1,2}=0,\quad A_0=A_3=-\frac{\pi K}{t^2}\delta(\omega -d_0-d_3).
\end{equation}
When $K$ is finite, this readily implies the surface state dispersion relation
\begin{equation}\label{eq:Esurface}           \varepsilon_\bk=d_0+d_3=\sum_{i=x,z}D_i (1-\cos{k_i})+t_1\sin{k_{x}},%-\mu,
\end{equation}
which mainly accounts for the circulating chiral states on the top surface, and is modified by the extra dispersions in the 2D surface BZ. As aforementioned and shown in Fig.~\ref{fig:SSreg-band_main}(B2), finite $D_z$ bends the Fermi arc towards $k_x$-direction to make it curved, otherwise the arc becomes a straight line. 
This is also consistent with earlier low-energy approximate solutions for simpler models\cite{Okugawa2014,Zhang2016a,Zhang2022}.
Furthermore, the fact that only $A_0,A_3$ are finite and identical implies a locking between the charge and spin-$S_z$ channels; we also find the surface state entirely polarized in spin-$\uparrow$. %as we will find $K<0$ below. 
Such a locking and spinful feature is directly visible in Fig.~\ref{Fig:Aomegakxkz_main}, where the surface band appears identically in the charge and $S_z$ channel.

The resonance at the surface state (henceforth denoted by index `ss') band \eqref{eq:Esurface} immediately leads to the relation $F_\mathrm{ss}=0,C_\mathrm{ss}=B_\mathrm{ss}^2$ and $K_\mathrm{ss}= B_\mathrm{ss}+r|B_\mathrm{ss}|$ with 
\begin{equation}\label{eq:Bss}
    B_\mathrm{ss}=d_{12}^2-t^2,
\end{equation}
where $C_\mathrm{ss}\equiv C|_{\omega=\varepsilon_\bk} \mycomment{C|_{\bar d_0=d_3}}$ and similarly for others.
Here, $C_\mathrm{ss}\geq0$ generally satisfies the assumption of \eqref{eq:specA} self-consistently except for the special case $C_\mathrm{ss}=0$.
Thus, $C_\mathrm{ss}=0$ defines a momentum-space borderline of the surface state, i.e., 
\begin{equation}\label{eq:borderline_k}
    B_\mathrm{ss}(\bk)=0,
\end{equation}
which forms a loop that touches the two WPs projected in the 2D BZ, as shown in Fig.~\ref{fig:SSreg-band_main}(B1). %The interesting behavior around the borderline will be discussed later. 
To have a nonvanishing spectral weight of the surface state in \eqref{eq:specA}, $K_\mathrm{ss}$ needs to be finite and negative such that $A_0>0$ as aforementioned.
Inside the borderline, i.e., within the constraint momentum-space region \begin{equation}\label{eq:constraint}
    B_\mathrm{ss}(\bk)<0,
\end{equation}
the $r=-1$ solution is chosen and one finds 
\begin{equation}\label{eq:A03ss_in}
    A_{0,3}^\mathrm{ss}(\omega,\bk)= 2\pi R(\bk)\delta(\omega-\varepsilon_\bk),
\end{equation}
where $R(\bk)=-B_\mathrm{ss}/t^2=(1-d_{12}^2/t^2)$.
Hence, \eqref{eq:constraint} gives the physically allowed region in the 2D momentum space, within which the surface band \eqref{eq:Esurface} is defined. 
Outside this region one has $B_\mathrm{ss}>0$ and the only nonvanishing choice $r=1$ leads to unphysical $A_0<0$, we thus still have $r=-1$ and hence
\begin{equation}\label{eq:A03ss_out}
    A_{0,3}^\mathrm{ss}=0,
\end{equation} 
which is physically expected beyond the surface state constraint and outside the bulk continuum.

\begin{figure*}[!htbp]
\centering
\includegraphics[width=17.8cm]{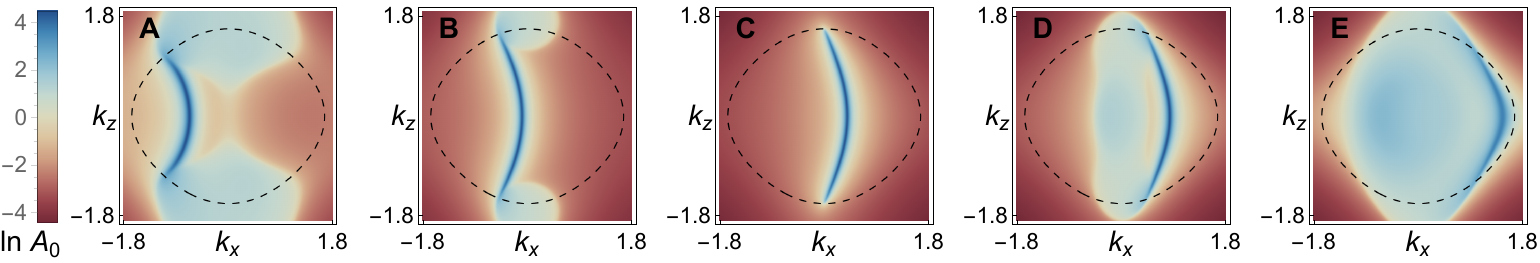}
\caption{ARPES signal log-scaled in the charge channel $A_{0}$ at several energy cuts showing the Fermi arc surface state (deep or dark blue) merging into the bulk. Panels (A-E), energies from low to high, correspond to the Fermi arcs as black lines in Fig.~\ref{fig:SSreg-band_main}(B2); panel (C) is the case when $E_F$ is pinned to the WP. Dashed lines indicate the boundary of the surface state constraint region, corresponding to Fig.~\ref{fig:SSreg-band_main}(B1). The momentum-dependent intensity profile and shape of the arc resonance encode complementary and key information about the WSM system. Parameters same as Fig.~\ref{Fig:Aomegakxkz_main}.
}\label{Fig:AkxkzArc_main}
\end{figure*}

It is intriguing to note that the momentum dependence of the resonance in \eqref{eq:A03ss_in} can be measured from the intensity $R(\bk)$ in both charge and spin channels of ARPES, which encodes the information of the surface state constraint function $B_\mathrm{ss}$ or the Hamiltonian components $d_{1,2}$. 
%Furthermore, an extra feature worth pointing out is that 
Such information is \textit{complementary} to the surface band dispersion \eqref{eq:Esurface}, which involves $d_{0,3}$ only and is measured through the surface band shape in ARPES as determined by the $\delta$-function in \eqref{eq:A03ss_in}. Therefore, from the resonance intensity $R(\bk)$ and the band shape $\varepsilon_\bk$, most information of the bulk WSM system %, if not all, 
can be extracted by merely inspecting surface resonance \eqref{eq:A03ss_in}, realizing an intriguing holography-like situation.
For instance, as shown in Fig.~\ref{Fig:AkxkzArc_main}, since ARPES measures photoemission intensity of occupied bands, at different chemical potentials tunable via gate voltage control, the Fermi arc in the 2D BZ traverses across the constraint region \eqref{eq:constraint}, and serves as a tomography of the surface resonance of the WSM. Here, we only exemplify the behavior of one particular surface. In experiments, different crystalline surfaces can be similarly measured, which will extract more complete information about the constraint and the Hamiltonian, including the anisotropy of the system. In other WSM systems where the surface carries more complex spin structures, in-plane $A_{1,2}$ can also supply further information.

In addition, with such a direct characterization of the surface state, we are able to point out yet another important but often overlooked phenomenon: the numbers of Fermi arcs and WP pairs are not necessarily identical. The common belief of a one-to-one correspondence between them assumes that an energy plane cuts a surface state continuum and gives an arc. 
In contrast, the constraint region \eqref{eq:constraint} may not be simply connected. As shown in Fig.~\ref{fig:SSreg-band_main}(C1,C2) for a minimal noncentrosymmetric model with two WP pairs (SI Sec.~\ref{SM:Pbreak}), it can generally be \textit{multiply connected} with %$n$ 
inner holes, then the isoenergy curve is disconnected once it passes a hole, i.e., a forbidden region, even if the Fermi energy $E_F$ is at WPs. This leads to more Fermi arcs for one WP pair, depending on $E_F$ and also the hole position. In experiments, discretion is necessary when identifying or counting WPs from arc spectroscopy; instead, knowledge of the constraint region would be helpful. %and observation of linear dispersion help confirmation. 
A hole, i.e., an inner boundary of the constraint region, is not fundamentally different from the usual outer boundary where arcs end; both are merging borderlines from surface to bulk and our later discussion applies equally well. 
For instance, an inner hole borderline can also pass WPs as the outer one does.
The above phenomenon of excess Fermi arcs is very general and can appear in many other WSMs. In fact, based on the present formalism applied to WSMs with multiple WP pairs, we can accurately analyze more generic but unconventional properties of the projected surface connection pattern (SI Sec.~\ref{SM:connectivity}). Strongly affected by the shape of constraint region and surface state energy surface and especially tuned by $E_F$, the naive association of a Fermi arc with a WP pair is often not the case.
%other factors of the constraint and dispersion can also affect the number of arcs

\subsection*{Signatures from bulk continuum}
Since Green's function \eqref{eq:ginv&g} has the bulk beneath the top surface integrated out, it should have the bulk state contribution included as well, albeit in an off-resonance manner. As aforementioned, away from the energy-momentum space region that admits possible surface state, we do not necessarily have $C\geq0$. Rewriting \eqref{eq:C} as $C=(\bar d_0^2-\cE_+^2)(\bar d_0^2-\cE_-^2)$ with $\cE_s=\sqrt{d_3^2+(d_{12}+s\,t)^2}$, we require $C<0$ and generally find two disconnected regions in the energy-momentum space for $s=\pm1$
\begin{equation}\label{eq:cE_t}
    \cE_{-s}^2 < (\omega-d_0)^2 < \cE_s^2.
\end{equation}
They exactly correspond to the projection of bulk states onto the surface, which manifests as conduction and valence bands. Within such continuum regions, $K$ of \eqref{eq:Ks} acquires a finite \textit{imaginary} part even before the analytic continuation and dominates the spectral function \eqref{eq:cA}; such a contribution is thus off-resonance, i.e., finite and not peaked as $\delta$-functions as the surface state. Therefore, $\cE_s$ should signify the band edges of the bulk states measured from $d_0$. We exemplify such band edges in Fig.~\ref{Fig:Aomegakxkz_main}(A1,B1). In fact, we note from Hamiltonian \eqref{eq:H_main} that the bulk eigenenergy (measured from $d_0$) squared reads
\begin{equation}
\begin{split}
    \cE^2(\bk)%=&(d_1-t\cos{k_y})^2+(d_2+t\sin{k_y})^2+d_3^2\\
    %=&d_3^2+(d_{12}^2-2td_1\cos{k_y}+2td_2\sin{k_y}+t^2)\\
    = d_{12}^2+2|t|d_{12}\sin{(k_y-\phi})+t^2+d_3^2,
\end{split}
\end{equation}
where $\phi$ is the polar angle of the coordinate $(td_2,td_1)$.
Its two extrema %at $k_y=0,\pi$ 
are exactly the foregoing $\cE_{\pm}^2$. When such band edges accidentally touch each other, the effective Green's function \eqref{eq:ginv&g} will become singular (SI Sec.~\ref{SM:g_singularity}).

It is noteworthy that the choice $r=\pm1$ of solution branches in \eqref{eq:ginv&g} becomes in general more complicated outside the surface resonance, including the bulk continuum and the merging borderline.
However, one can reach an exceptionally simple criterion at \textit{any} $\bk$ and $\omega$, not limited to surface or bulk contribution (\textit{Methods}): find the choice of $r$ that maximizes $A_0(r)$. This is how all the spectral functions are evaluated throughout this study.
%\footnote{An alternative and equivalent way is to find the first $A_0(r)>0$ in the order of $s=-1$ and $s=1$.}. 

In Fig.~\ref{Fig:Aomegakxkz_main}, we indeed observe the projection of bulk conduction and valence bands in the surface BZ and they are connected through the surface band in between. 
In Fig.~\ref{Fig:AkxkzArc_main}(A-C), the Fermi arc similarly connects two bulk Weyl cones separated along $k_z$ direction; the bulk state is itself connected at energy higher than the WP in Fig.~\ref{Fig:AkxkzArc_main}(D,E), which is due to the distortion from $d_0$ and consistent with Fig.~\ref{Fig:Aomegakxkz_main}(B1).
Also, as aforementioned, the spectral intensity of the projected bulk bands remains independent from $\eta$ in \eqref{eq:A_alpha}, in contrast to the surface resonance.
From \eqref{eq:ginv&g} and \eqref{eq:A_alpha}, we know that $A_{1,2}$ and $A_0-A_3$ are finite \textit{only} when $C<0$ for the bulk continuum region, so they all vanish for the surface state resonance as discussed previously. An immediate implication is that the in-plane spin spectral weight $A_{1,2}$ only appears in the bulk continuum and respectively follows the profile of $d_{1,2}$, e.g., both $d_2$ and $A_2$ changes sign at $k_x=\pm\pi/2$, as we observe in Fig.~\ref{Fig:Aomegakxkz_main}.  
%For instance, since $d_2$ in the present case changes sign at $k_x=\pm\pi/2$, so does $A_2$ in Fig.~\ref{Fig:Aomegakxkz_main}(A3).
$d_3=t_1\sin{k_x}$ seemingly suggests $A_3(\omega,k_x)=-A_3(\omega,-k_x)$, which does not accurately hold in Fig.~\ref{Fig:Aomegakxkz_main}(A4) although the trend remains in gross. This is because such bulk symmetry, under surface projection, will be distorted, as exemplified by the chiral surface state obviously asymmetric for $k_x$. In fact, if the opposite surface was included, the overall symmetry could be restored for the whole sample.

\subsection*{Singular behavior along merging borderline}
\begin{figure*}[!htbp]
\centering
\includegraphics[width=16cm]{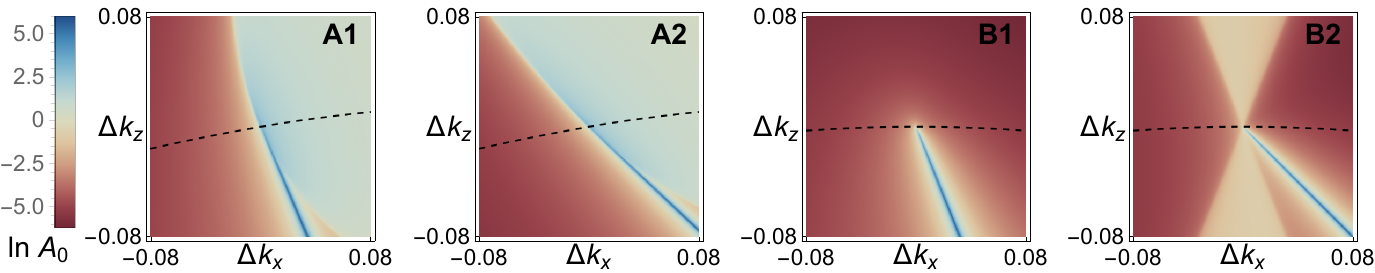}
\caption{ARPES signal $A_{0}(\Domega=0,\Dbk)$ based on \eqref{eq:A03asymptote_blAll_main} log-scaled in the $\bk$-space vicinity of a borderline point at the origin: (A1,A2) a generic borderline point with $k_x=-0.2$; (B1,B2) the WP at $k_z=k_w$. Dashed lines indicate the upper boundary of the surface state constraint region as introduced in Fig.~\ref{fig:SSreg-band_main}(B1). (A1,B1) Type-I case with parameters same as Fig.~\ref{Fig:Aomegakxkz_main} and respectively corresponding to the upper merging point in Fig.~\ref{Fig:AkxkzArc_main}(B,C); (A2,B2) type-II case for the same model with larger $D_z=1$. 
(A) Deep blue Fermi arc surface state merges into the bulk continuum but is anomalously weakened albeit still divergent upon merging [see Fig.~\ref{Fig:Aomegakxkz3D_main}(A)], which is hence less discernible partially due to the log scale. (B) Fermi arc ends at WP where the signal is finite [see Fig.~\ref{Fig:Aomegakxkz3D_main}(B)]; (B2) type-II case shows Fermi pockets connected to WP.
}\label{Fig:Adkxdkzbl_main}
\end{figure*}

We inspect how the surface state merges into the bulk and the behavior around the merging points, i.e., the $C=0$ borderline, which is a twofold restriction in the energy-momentum space as shown in Fig.~\ref{fig:SSreg-band_main}(B,C): the loop-like border of the surface state constraint \eqref{eq:borderline_k} in the surface BZ and the energy pinned to the corresponding surface band. 
Since $C_\mathrm{ss}=B_\mathrm{ss}^2>0$ except for the borderline, the entire surface band \eqref{eq:Esurface} at \textit{any} $\bk$ in the whole surface BZ (dashed line in Fig.~\ref{Fig:Aomegakxkz_main}(A1,B1)), not limited to the constraint region \eqref{eq:constraint}, is separated from the bulk states ($C<0$). Hence, they will tangentially touch each other at the borderline. The real surface state restricted by the constraint thus merges into the bulk continuum at the borderline, as shown in Fig.~\ref{Fig:Aomegakxkz_main} and Fig.~\ref{Fig:AkxkzArc_main}.
Although the surface resonance is outside the bulk continuum, its residual influence, according to \eqref{eq:ginv&g} and \eqref{eq:A_alpha}, enters $A_0$ and $A_3$ of the bulk continuum through the energy denominator (index `bs' for bulk states)
\begin{equation}\label{eq:A_bs}
    (A_0+A_3)^\mathrm{bs}=\frac{2\Im K}{t^2(\omega-\varepsilon_\bk)}.
\end{equation}
%while $(A_0-A_3)^\mathrm{bs}$ is suppressed by its numerator with $(\omega-\varepsilon_\bk)$.
This means stronger spectral weight in $A_0$ and $A_3$ if close to the \textit{full} surface state dispersion $\varepsilon_\bk$ of \eqref{eq:Esurface} regardless of the constraint, as seen in Fig.~\ref{Fig:Aomegakxkz_main}(A1,A4,B1,B4)\mycomment{\red{...}}. This is physically understandable as the resonant surface spectral weight will not suddenly disappear but smear into the bulk continuum. 
%We use index `bl' to denote quantities evaluated at the borderline with $C_\mathrm{ss}=0$.

\begin{figure*}[!htbp]
\centering
\includegraphics[width=17.8cm]{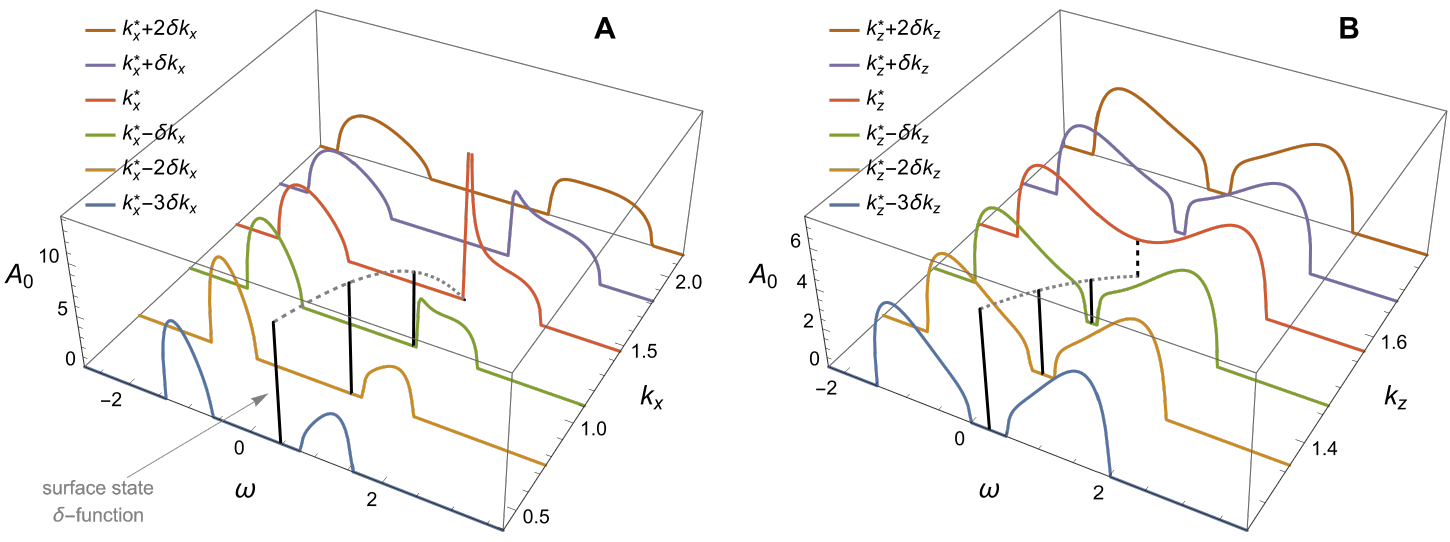}
\caption{Evolution and transition of ARPES signal $A_{0}(\omega)$ in the vicinity of the merging borderline: (A) a generic borderline point at $k_x^*$ corresponding to the upper right merging point in Fig.~\ref{Fig:Aomegakxkz_main}(A); (B) the WP at $k_z^*=k_w$.
Respectively along the dotted horizontal and vertical paths in Fig.~\ref{fig:SSreg-band_main}(B1), panel (A) selects successive $k_x$ cuts with an incremental step $\delta k_x=0.35$ and panel (B) selects similarly along $k_z$ with $\delta k_z=0.1$. Vertical solid lines indicate the surface state resonance with the height proportional to the intensity of $\delta$-function; their intensity envelope function determined by $R(\bk)$ is shown as a grey dashed curve vanishing at the borderline point, which is passed by the red signal. From blue to red and to brown signals: (A) The $\delta$-function peak is weakened as $\propto|\Dbk|$ and eventually merges with the bulk signal to produce the red anomaly as $A_0  \propto | \Domega |^{-1/2} $ at the borderline, after which only bulk state signal is left; (B) The $\delta$-function peak is weakened while the bulk gap shrinks till the WP, where they touch each other to produce a finite signal indicated by the vertical dashed line, after which the bulk gap opens again but without surface state resonance.
}\label{Fig:Aomegakxkz3D_main}
\end{figure*}

Approaching the merging borderline from the bulk continuum, both the nominator and denominator reduce to zero in \eqref{eq:A_bs}, which becomes apparently \textit{undetermined}. The similar happens in \eqref{eq:specA} or \eqref{eq:A03ss_in} from the surface resonance side as key quantities $C,B,K$ all approach zero. To reveal the singular behavior thereof for \textit{any} approaching direction, we consider the expansion around the momentum $\bk_\mathrm{bl}$ and energy $\varepsilon_{\bk_\mathrm{bl}}$ on the borderline respectively (denoted by index `bl'), i.e.,
$\Dbk=\bk-\bk_\mathrm{bl},\Domega=\omega-\varepsilon_{\bk_\mathrm{bl}}$. 
One can cast quantities in \eqref{eq:Ks} up to the relevant orders in $\Dbk$ and $\Domega$
\begin{equation}\label{eq:BC_expand_general_main}
\begin{split}
    B&=B_\mathrm{ss}+F\approx\balpha\cdot\Dbk+F_1\\%=\balpha'\cdot\Dbk-2d_{3}(\bk_\mathrm{bl})\Domega_\eta\\
    C&\approx 4t^2 F_1-c(\Domega_\eta)^2+\Dbk\cdot\Lambda\cdot\Dbk + \Domega_\eta\bgamma\cdot\Dbk    
\end{split}
\end{equation}
where all parameters like $\balpha=\partial_\bk B_\mathrm{ss}\vert_\mathrm{bl}$ and rank-2 tensor $\Lambda$ are evaluated along the borderline and given in \textit{Methods}. $F_1= -2 d_{3}(\bk_\mathrm{bl})f$ is the linear order term of $F$, and with notation $\Domega_\eta=\Domega+\ii\eta$, we denote 
\begin{equation}\label{eq:f_main}
    f=\Domega_\eta-\partial_\bk\varepsilon\vert_{\bk_\mathrm{bl}}\cdot\Dbk.%,\quad f'=f+\partial_\bk d_{3}(\bk_\mathrm{bl})\cdot\Dbk
\end{equation}
While $F_1$ accounts for the leading linear effect in $C$ along almost the entire borderline, it vanishes at WPs as per the general model construction, hence the rest of quadratic terms will become important.  
With \eqref{eq:BC_expand_general_main}, the general spectral function around the entire borderline is
\begin{equation}\label{eq:A03asymptote_blAll_main}
\begin{split}
    A_{0,3}(\Domega,\Dbk)\approx  \Im\frac{B+r\sqrt{C}}{t^2f}.
\end{split}
\end{equation}
We visualize $\Domega=0$ case for representative borderline points away from and at the WP in Fig.~\ref{Fig:Adkxdkzbl_main}, where a type-II WSM case is exemplified in comparison when the velocity tilting is large enough to induce the Lifshitz transition\cite{Soluyanov2015} (SI Sec.~\ref{SM:typeII}). 
\mycomment{\red{......}}In Fig.~\ref{Fig:Aomegakxkz3D_main}, we additionally visualize the evolution and transition in the proximity of the merging borderline, in which the anomalous behavior across the merging along momentum axes manifests conspicuously.

We first approach the borderline along the surface state resonance. Following the earlier discussion of surface resonance spectroscopy, this sets $F=0$ in the first place in \eqref{eq:BC_expand_general_main} and leads to $B_\mathrm{ss}\approx\balpha\cdot\Dbk,C_\mathrm{ss}\approx B_\mathrm{ss}^2$. %Consistent with \eqref{eq:Bss} and \eqref{eq:A03ss_in}, n
Now \eqref{eq:A03asymptote_blAll_main} becomes in the limit $\eta\rightarrow0^+$
\begin{equation}\label{eq:A03asymptote_blss}
\begin{split}
    A_{0,3}(\Domega,\Dbk)\approx  \mycomment{\Im\frac{B_\mathrm{ss}}{t^2f}=}-\frac{2\pi\balpha\cdot\Dbk}{t^2}\delta(\Domega-\partial_\bk\varepsilon\vert_{\bk_\mathrm{bl}}\cdot\Dbk).
\end{split}
\end{equation}
The surface state constraint \eqref{eq:constraint} also applies and $\balpha$ defines the outward normal direction at the constraint region boundary, which is generally nonzero. Hence, its resonance intensity scales with the momentum normal to the borderline; $\balpha\cdot\Dbk<0$ exactly means that $\Dbk$ needs to be inward the constraint region to have finite signals.

\eqref{eq:A03asymptote_blss} misses the continuum side and the generic off-resonance behavior outside the bulk continuum. %, which are otherwise encoded in \eqref{eq:A03asymptote_blAll_main}. 
With deviation from the surface state resonance, the full form \eqref{eq:A03asymptote_blAll_main} captures the correct behavior as shown in Fig.~\ref{Fig:Adkxdkzbl_main}, where especially $F_1$ in $C$ becomes the leading linear contribution away from WPs.
If only this leading effect is concerned, the formula reduces to
\begin{equation}\label{eq:A03asymptote_bl_main}
\begin{split}
    A_{0,3}(\Domega,\Dbk)\approx \Im\frac{r\sqrt{C}}{t^2f}\approx \frac{2r}{|t|}\Im\sqrt{\frac{-2 d_{3}(\bk_\mathrm{bl})}{f}}.
\end{split}
\end{equation} 
%depends on the borderline momentum $\bk_\mathrm{bl}$ through $d_{3}(\bk_\mathrm{bl})$ and $\partial_\bk\varepsilon|_\mathrm{bl}$ in $f$.
This formula determines the local merging configuration, including whether the borderline is approached from the bulk continuum or the surface state region near the conduction or the valence band (see \textit{Methods}).
Besides, along the borderline it exhibits a singularity scaled as $A_{0,3}^\mathrm{bl}\propto |\Domega-\partial_\bk\varepsilon\vert_{\bk_\mathrm{bl}}\cdot\Dbk|^{-\frac{1}{2}}$ with $\Domega,\Dbk\rightarrow0$. Without much loss of generality, we will refer to it by $|\Domega|^{-\frac{1}{2}}$-singularity.
Therefore remarkably, the borderline is a set of singular \textit{branch points} with unconventional inverse square root divergence $|\Domega|^{-\frac{1}{2}}$, in contrast to the usual $\delta$-function due to a \textit{simple pole} divergence $|\Domega|^{-1}$.
It resolves the foregoing undeterminedness %, including that of \eqref{eq:A03asymptote_blss}, 
in any direction, where the spectral weight changes between resonant $\delta$-function divergence (surface-state side), $|\Domega|^{-\frac{1}{2}}$-singularity (borderline), and finite (bulk side).
Accordingly in Fig.~\ref{Fig:Aomegakxkz3D_main}(A), the blue, yellow, and green signals approach the merging point (red signal) from the surface side while the associated surface state resonance is weakened and transitions to the anomalously peaked resonance at merging; further into the bulk side, such a $|\Domega|^{-\frac{1}{2}}$-divergence changes to the smooth purple and brown signals. This elucidates what happens in Fig.~\ref{Fig:AkxkzArc_main}(A,B) and Fig.~\ref{Fig:Adkxdkzbl_main}(A) near the merging. 
The borderline forms a set of critical transition points of the spectral signal in ARPES, regardless of the approaching direction in energy-momentum space. With finite relaxation strength $\eta$, the signal variation could become a smoother crossover, but the scaling behavior, including that of the linewidth,
bears the same transition. 

Around WPs, $C$ is fully quadratic in deviations and \eqref{eq:A03asymptote_blAll_main} takes the form
\begin{equation}\label{eq:A03asymptote_wp_main}
\begin{split}
    &A_{0,3}(\Domega,\Dbk)\\
    \approx \,& \Im\frac{\balpha\cdot\Dbk+r\sqrt{-c(\Domega_\eta)^2+\Dbk\cdot\Lambda\cdot\Dbk + \Domega_\eta\bgamma\cdot\Dbk}}{t^2f},
\end{split}
\end{equation}
where $\balpha=\partial_\bk B_\mathrm{ss}\vert_\mathrm{wp},c=4t^2,\bgamma=8t^2\partial_\bk d_0\vert_\mathrm{wp}$ and $\Lambda=\balpha\balpha+4t^2(\partial_\bk d_3\partial_\bk d_3-\partial_\bk d_0\partial_\bk d_0 )_\mathrm{wp}$.
This formula fully captures the merging configuration around WPs, especially the distinction between type-I and type-II cases, as shown in Fig.~\ref{Fig:Adkxdkzbl_main}(B), is directly related to the principal values of the tensor $\Lambda$ (see \textit{Methods}).
Approaching the WPs, \eqref{eq:A03asymptote_wp_main} gives $A_{0,3}^\mathrm{wp}=  2/|t|$ with $r=-1$ chosen. This singles out the WP as a unique \textit{removable singularity} of spectral weight as shown in Fig.~\ref{Fig:Adkxdkzbl_main}(B), which purely depends on the hopping amplitude and does not diverge with vanishing $\eta$. 
In Fig.~\ref{Fig:Aomegakxkz3D_main}(B), we observe that the bulk gap closes smoothly and reopens; more specifically, it shrinks to zero while the intensity of surface state resonance is weakened to zero upon reaching the merging point (WP), where it transitions to the red finite WP signal.
%further into the bulk side, gapped bulk signal recurs with no resonance any more.
The surface states entirely end at WPs and WPs simultaneously belong to the bulk continuum; this physically explains why the WP signal eventually does not diverge as surface states do.
It thus is distinct from the foregoing surface-borderline-bulk transition and instead admits either a surface-bulk transition as in Fig.~\ref{Fig:Aomegakxkz3D_main}(B) (from $\delta$-function divergence to finite) or a borderline-bulk transition (from $|\Domega|^{-\frac{1}{2}}$-divergence to finite), depending on whether the WP is approached from the majority of surface states or along the borderline.
Note that the intensity of  $|\Domega|^{-\frac{1}{2}}$-singularity decreases with the distance $|\Dbk|$ from the WP
along the borderline.

These discussions reveal the strong inhomogeneity along the borderline in terms of both the near-merging behavior and the singularities themselves, reflecting a fundamental aspect of the topology generated between WPs in a WSM.
While the borderline can be viewed as a special edge curve of the surface state where distinct phenomena occur, there also exist WPs as the special points on the borderline.
The surface/bulk distinction plays an important role and one can approach the borderline from either side. On each side, it exhibits nontrivial energy-dependence and highly anisotropic angle $\arg\Dbk$-dependence for the \textit{entire} borderline. 
Such information provides important profiling of the WSM system, as reflected in Fig.~\ref{Fig:Adkxdkzbl_main} and Fig.~\ref{Fig:Aomegakxkz3D_main}, and can be measured in ARPES directly, e.g., through inspecting the endpoint behavior of Fermi arcs in Fig.~\ref{Fig:AkxkzArc_main}. %Similar information is also included in Fig.~\ref{Fig:Aomegakxkz3D_main}.

\subsection*{STS measurement}\label{Sec:STS}

%Images must be provided at final size, preferably 1 column width (8.7cm). Figures wider than 1 column should be sized to 11.4cm or 17.8cm wide. Numbers, letters, and symbols should be no smaller than 6 points (2mm) and no larger than 12 points (6mm) after reduction and must be consistent. 
\begin{figure*}[!htbp]
\centering
\includegraphics[width=17.4cm]{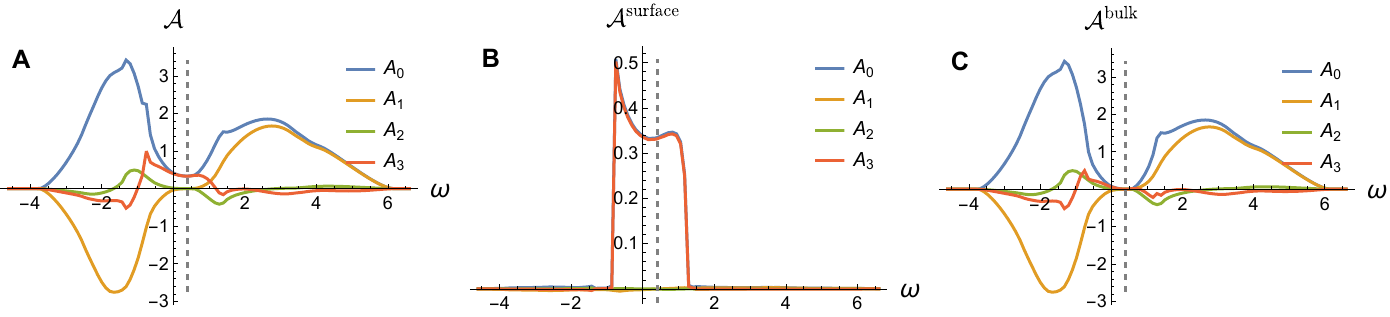}
\caption{Spin-resolved STS signals as a function of frequency. Integrated spectral functions $A_{0,1,2,3}$ in the charge and spin channels are shown in panel (A). Panels (B,C) respectively separate out the topological surface state contribution and the bulk contribution in (A). Dashed line indicates the WP energy. Parameters same as Fig.~\ref{Fig:Aomegakxkz_main}.
}\label{Fig:STS_main}
\end{figure*}

STS measurement finds differential conductance from varying the bias voltage of the (spin-polarized) tip. It is able to reconstruct the local density of the electronic states and is directly related to the spectral function \eqref{eq:A_alpha} integrated over momentum, i.e., $\cA_\alpha(\omega)=S_\mathrm{BZ}^{-1}\int\dd\bk A_\alpha(\omega,\bk)$ with $S_\mathrm{BZ}$ the area of the surface BZ. An extra merit of STS is that it can probe both occupied and empty states by simply changing the sign of bias voltage. This is, however, achievable in ARPES by directly tuning the chemical potential or with pump-probe ARPES\cite{Belopolski2017}. In the presence of impurities, quasiparticle interference patterns of STS related to spatially dependent $\cA(\omega,\br)$ can also detect the surface Fermi contour\cite{Chang2016,Inoue2016,Deng2016,Zheng2016,Yuan2019}.

Fig.~\ref{Fig:STS_main} shows the integrated spectral weight in charge and spin channels. The asymmetry with respect to the WP energy is due to the broken particle-hole symmetry in \eqref{eq:H_main}. By virtue of our analytical solution that physically identifies the meaning of the key quantity $C$, the surface and bulk contributions can also be clearly separated as $\cA^\mathrm{surface}$ and $\cA^\mathrm{bulk}$. Corroborating with our earlier discussion, only $\cA_0^\mathrm{surface}$ and $\cA_3^\mathrm{surface}$ are finite while $\cA^\mathrm{bulk}$ comes from all channels. 
The van Hove singularity in $\cA^\mathrm{surface}$ originates from the lowest energy part of the surface band in Fig.~\ref{fig:SSreg-band_main}(B2), which can become relatively flat as the band minimum. This particularly happens when the Weyl vector, the separation vector between WPs, is long such that the constraint region covers the surface band minimum. Therefore, it can be used as a phenomenon for estimating the Weyl vector scale important in experiments.
Near the WP energy, bulk spectral weight vanishes in every channel while that from the surface state dominantly contributes in a way that manifests the locking between the charge and spin-$S_z$ channels, which can be used as direct experimental evidence for identifying topological arc states. 
Besides directly accessing charge and spin channels in either ARPES or STS, one can also measure different surfaces of the sample that may bear arc states or not, from which the unique surface contribution can be clearly identified in comparison and even separated if the anisotropy between those surfaces is not too large.

\section*{Discussion and Summary}
The main WSM model we focused on can be realized in its own right in minimal magnetic Weyl materials and particularly also captures the surface spin polarization effect directly observable in spin-resolved ARPES and STS. 
Such discussion singles out the most important physical properties in general magnetic or noncentrosymmetric Weyl materials even with multiple pairs of WPs, since the pairwise connection of projected WPs on a surface is a common and distinguishing feature.  
In fact, the formalism is general enough to account for more interesting cases, e.g., the time-reversal-breaking WSM with more WP pairs and the inversion-breaking WSM with more pairs such as the one shown in Fig.~\ref{fig:SSreg-band_main}(C1,C2) and WPs not at the same energy (SI Secs.~\ref{SM:Pbreak} and \ref{SM:Tbreak}). 
Importantly, the main features discussed remain intact, despite the distinct shape of the surface state.
For noncentrosymmetric Weyl materials, the number of WP pairs has been reduced from twelve in TaAs to four\cite{Soluyanov2015,Ruan2016,Deng2016,Jiang2017} and the minimal two\cite{Koepernik2016,Belopolski2017}; increasingly many magnetic Weyl materials also appear in experiments and theoretical predictions, even with two or the minimal one pair of WPs\cite{Nie2017,Jin2017,WeylDiracReview,Nagaosa_2020,Lv2021}.

As we have already exemplified, all the analysis equally applies to type-II WSMs and the formulae and conclusions are fully general (SI Sec.~\ref{SM:typeII}). For example, an immediate observation is that although the surface band dispersion \eqref{eq:Esurface} will be strongly tilted by $d_0$, the surface state constraint region \eqref{eq:constraint} remains intact. 
Last but not least, the Dirac semimetal case shares similar physics as the signal will be effectively contributed by, e.g., one more pair of WPs, which can be readily captured by adding an extra chirality-reversed copy of the model in our discussion and none of the essential conclusions will change.

Based on a formalism of exactly solved WSM models at the surface, we present a unifying treatment of the topological surface state and the bulk continuum and reveal the crucial hybridization aspect of the bulk-boundary correspondence. This strongly broadens the physical access and enables a comprehensive analysis of the spectroscopic effects of the surface inseparable from the bulk. Important but often overlooked information about the bulk Hamiltonian and surface constraint can be revealed by surface spectroscopy. Unconventional transitions at the surface-bulk merging borderline are found via a systematic analysis of the highly anisotropic singular responses in the energy-momentum space. 
The fully analytical account and the general prediction will deepen our understanding of topological semimetals from their 3D surface-bulk hybrid nature and push forward the frontier of spectroscopy experiments on topological systems.

\section*{Methods}
\subsection*{Solution method of model Hamiltonian}
Here we sketch the method of solving the surface Green's function. 
For a sample semi-infinite in the $-\hat{y}$-direction, the Hamiltonian \eqref{eq:H_main} can be given in a \textit{recursive} form.
Contiguous $y$-layers start from the top surface and serve as the matrix indices $1,\cdots,\infty$
\begin{equation}\label{eq:H_recursive}
    H(k_x,k_z)=\begin{bmatrix}
        h & \cV \\
        \cV^\dag & H
    \end{bmatrix},
\end{equation}
where $h(k_x,k_z)$, given in the main text, is the Hamiltonian of the top layer decoupled from the bulk.
Also, the inter-$y$-layer coupling takes the form $\cV=(V,0, 0,\cdots)$ with $V=-t\sigma_-$ and $\sigma_\pm=\frac{1}{2}(\sigma_1\pm\ii\sigma_2)$.
We focus on the Green's function $g(\bk,z)=G_{11}$ of the top $y$-surface with generic complex frequency $z$ and $G=(z-H)^{-1}$ the original Green's function of $H$. Making use of the recursive property of \eqref{eq:H_recursive}, the self-energy of $g$ for the top surface is given by 
\begin{equation}
    \Sigma=\cV G \cV^\dag=V g V^\dag.
\end{equation}
When we have the bare Green's function of $h$ as $g_0=(z-h)^{-1}=(\bar d_0\sigma_0-\bd\cdot\bsigma)^{-1}$, the exact Dyson equation reads
\begin{equation}\label{eq:Dyson}
    g^{-1}=g_0^{-1}-\Sigma,
\end{equation}
which is an equation of $b_0,\bb$ in the effective surface Green's function $g=b_0\sigma_0+\bb\cdot\bsigma$.
Intuitively, the foregoing form of interlayer coupling $V$ given by $\sigma_\pm$ hides the information of $d_{1,2}$, which is `squeezed' but encoded in the surface $g$.
Hence, one finds the intriguing structure of \eqref{eq:ginv&g}, especially the importance of $d_{0,3}$ that appears in the energy denominator only for $b_0+b_3$.
In SI Sec.~\ref{SM:Hamiltonian}, we discuss in more detail how such a formalism can be applied to more general Hamiltonians.

%\subsection*{Choice of solution branch}
We in general face the issue of the choice of $r=\pm1$ in calculating \eqref{eq:cA}, which is now extra complicated by the inevitable branch cut of square root in \eqref{eq:Ks}. Firstly, it is a convenient choice to place the branch cut at $\arg{C}=\pi$; to account properly for the surface state case of $C>0$, $\arg{C}=0$ should be avoided, otherwise, it would cause an unphysical exchange between the shaded surface state constraint region in Fig.~\ref{fig:SSreg-band_main}(B1) and its complement. Secondly, as aforementioned, $r$ can be fixed by the positive-definiteness of the charge channel spectral weight $A_0$. Since the off-resonance bulk contribution is purely coming from the imaginary part $\Im K$, i.e., $r\sqrt{C}$ in \eqref{eq:Ks}, \textit{only one} $r$ makes $A_0>0$ and hence naturally determines the choice. Slightly away from the surface state resonance and when the analytic continuation in \eqref{eq:A_alpha} practically uses a finite $\eta$, such uniqueness of $r$ for positive $A_0$ may not hold everywhere, e.g., around the borderline of the constraint region where the surface state merges into the bulk states; this, however, can be resolved by noting that the spectral function varies smoothly away from the exact surface state resonance in \eqref{eq:A03ss_in}. 
With these considerations combined, we reach the simple but general criterion at \textit{any} $\bk$ and $\omega$: find the choice of $r$ that maximizes $A_0(r)$.

\subsection*{Surface state resonance and behavior near the borderline}
According to \eqref{eq:ginv&g}, the effect of surface state resonance can only enter the component below of the spectral function via the substitution $z\rightarrow\omega+\ii\eta$
\begin{equation}\label{eq:A03_ss}
\begin{split}
    (A_0+A_3)^\mathrm{ss}%=&-4\lim_{\eta\rightarrow0^+}\Im\frac{-K}{2t^2(\omega+\ii\eta -\varepsilon_\bk)}\Big|_\mathrm{ss}\\
    =&\lim_{\eta\rightarrow0^+}\frac{(\omega-\varepsilon_\bk)\Im K_\mathrm{ss}-\eta\Re K_\mathrm{ss}}{t^2[(\omega-\varepsilon_\bk)^2+\eta^2]/2}\Big|_{\omega\rightarrow\varepsilon_\bk},
\end{split}
\end{equation}
where only the second part with $\Re K_\mathrm{ss}$ matters and gives the Lorentzian structure, because $\lim_{\eta\rightarrow0^+}\Im K_\mathrm{ss}=0$ outside the bulk continuum.
We thus target the charge-spin-locked $A_0=A_3$ in the following. 
In the general $K_\mathrm{ss}(\omega+\ii\eta)$ on surface state resonance, we use \eqref{eq:Bss} and separate out the contribution due to relaxation $\eta$ 
\begin{equation}\label{eq:Kss0}
\begin{split}
    K_\mathrm{ss}(\omega+\ii\eta) &= B_\mathrm{ss} + \kappa + r\sqrt{B_\mathrm{ss}^2+\kappa[2(B_\mathrm{ss}+2t^2)+\kappa]} 
\end{split}
\end{equation}
where $\kappa(\omega)=\eta(\eta-2\ii\bar d_0)$. % and $B_\mathrm{ss}=d_{12}^2-t^2$. 
When $B_\mathrm{ss}\neq0$, i.e., away from the surface-bulk borderline, we have 
\begin{equation}\label{eq:Kss1}
\begin{split}
    K_\mathrm{ss}(\omega+\ii\eta) %&\approx B_\mathrm{ss} + \kappa + r\sqrt{B_\mathrm{ss}^2+2\kappa(B_\mathrm{ss}+2t^2)} \\
    &\approx (B_\mathrm{ss}+r|B_\mathrm{ss}|) + \kappa (1+r\frac{B_\mathrm{ss}+2t^2}{|B_\mathrm{ss}|}),
\end{split}
\end{equation}
where we retain the leading contribution of $\eta$. Inside the surface state constraint region \eqref{eq:constraint}, $(A_0+A_3)^\mathrm{ss}$ is contributed only by $B_\mathrm{ss}+r|B_\mathrm{ss}|$ because the second part in the second line of \eqref{eq:Kss1} is at least $\sim\eta^2$ in its real part; this indicates the choice $r=-1$ aforementioned and we obtain \eqref{eq:A03ss_in} in the main text, where $\delta(\omega-\varepsilon_\bk)$ is from the resonance $\propto\eta^{-1}$ suggested in \eqref{eq:A03_ss}.

%\subsection*{Behavior near the borderline}
To study the property in the vicinity of the borderline between the surface state and the bulk, it suffices to focus on $A_{0,3}$ according to \eqref{eq:ginv&g}.
Importantly, the full effect up to the quadratic order in $\Domega,\Dbk$ is relevant to our discussion.
We separate $B=B_\mathrm{ss}+F$ using \eqref{eq:Bss}. Up to the leading order, we firstly find $B_\mathrm{ss}\approx\balpha\cdot\Dbk$, where $\balpha=\partial_\bk B_\mathrm{ss}\vert_\mathrm{bl}=2\sum_{i=1,2}d_i\partial_\bk d_i\vert_\mathrm{bl}$, dependent on $d_{1,2}$ that determines the constraint region \eqref{eq:constraint}, defines the normal direction at the boundary of the constraint region and is in general finite along the entire borderline. %$B_\mathrm{bl}$ is $B_\mathrm{ss}$ restricted to $\bk_\mathrm{bl}$. 
Secondly, we have $F\approx -\bar d_0^2+ d_3^2 +\partial_\bk F\cdot\Dbk+ \frac{1}{2}\Dbk\cdot\partial_\bk^2 F\cdot\Dbk \vert_\mathrm{bl}=F_1+F_2$ with
\begin{equation}\label{eq:F10}
\begin{split}
    F_1=&-2 d_{3}(\bk_\mathrm{bl})\Domega_\eta+  \bbeta\cdot\Dbk 
    %=& -2 d_{3}(\bk_\mathrm{bl})f-2\ii\eta f'\\
    =  -2 d_{3}(\bk_\mathrm{bl})f\\
    F_2=&-(\Domega_\eta)^2 + \Dbk\cdot\Gamma'\cdot\Dbk + \frac{1}{c}\Domega_\eta\bgamma\cdot\Dbk,
\end{split}
\end{equation}
where $\bbeta=\partial_\bk F\vert_\mathrm{bl}= 2d_{3}\partial_\bk\varepsilon|_\mathrm{bl}$, $\bgamma=8t^2\partial_\bk d_0\vert_\mathrm{bl}$, $c=4t^2$, and the rank-2 Hessian tensor $\Gamma'=\frac{1}{2}\partial_\bk^2 F\vert_\mathrm{bl}=\Gamma +d_3\partial_\bk^2\varepsilon\vert_\mathrm{bl}$ with $\Gamma=\partial_\bk d_3\partial_\bk d_3-\partial_\bk d_0\partial_\bk d_0\vert_\mathrm{bl}$. 
Then we have 
\begin{equation}\label{eq:B_expand_general}
    B\approx\balpha\cdot\Dbk+F_1=\balpha'\cdot\Dbk-2d_{3}(\bk_\mathrm{bl})\Domega_\eta
\end{equation}
up to the linear order with $\balpha'=\balpha+2d_3\partial_\bk\varepsilon\vert_\mathrm{bl}$, which suffices for our purpose. This leads to 
\begin{equation}\label{eq:C_expand_general}
    C\approx cF_1-c'(\Domega_\eta)^2+\Dbk\cdot\Lambda'\cdot\Dbk + \Domega_\eta\bgamma'\cdot\Dbk
\end{equation}
with $c'=c-4{d_{3}(\bk_\mathrm{bl})}^2$, $\Lambda'=c\Gamma'+\balpha'\balpha'$, and $\bgamma'=\bgamma-4d_{3}(\bk_\mathrm{bl})\balpha'$. 
In the vicinity of WPs, \eqref{eq:B_expand_general} and \eqref{eq:C_expand_general} become
\begin{equation}\label{eq:BC_expand_WP}
\begin{split}
    B_\mathrm{wp}&\approx\balpha\cdot\Dbk \\
    C_\mathrm{wp}&\approx -c(\Domega_\eta)^2+\Dbk\cdot\Lambda\cdot\Dbk + \Domega_\eta\bgamma\cdot\Dbk    
\end{split}
\end{equation}
with $\Lambda=c\Gamma+\balpha\balpha$ and all parameters evaluated at the WPs. Note that we use primed notation explicitly for parameters defined for generic borderline points, which is omitted in \eqref{eq:BC_expand_general_main} for brevity. With these definitions, \eqref{eq:ginv&g} gives
\begin{equation}
\begin{split}
    A_0-A_3\approx  \Im\frac{2f(K+2t^2)}{t^2(t^2+\balpha\cdot\Dbk)} \approx\Im\frac{2(fK+2\ii\eta t^2)}{t^4}.
\end{split}
\end{equation}
Besides the unimportant $\eta$-linear term, its leading order is generally nonlinear in $\Domega,\Dbk$ and higher order than $K$, which is negligible compared to $A_0-A_3$ below. Hence, we have identical $A_0,A_3$ within the leading order around the entire borderline and can obtain \eqref{eq:A03asymptote_blAll_main} and what follows.

\subsection*{Determine the merging configuration}
Given the crucial role of $C$ in \eqref{eq:A03asymptote_blAll_main}, \eqref{eq:A03asymptote_bl_main}, and \eqref{eq:A03asymptote_wp_main}, %in the above formulae, 
it is clarifying to understand its physical meaning, which exactly determines the local merging configuration.
Along the borderline except WPs, regarding \eqref{eq:A03asymptote_bl_main}, $C\approx4t^2F_1$ is transparent per the surface/bulk distinction of $C\gtrless0$: changing the signs of $\Domega,\Dbk$, i.e., deviating in opposite directions away from the borderline, switches between surface and bulk contributions. 
For instance, setting $\Dbk=0$, one finds $C=-8t^2d_{3}(\bk_\mathrm{bl})\Domega$. Since $d_3$ is the chiral part of the surface band \eqref{eq:Esurface}, $d_{3}(\bk_\mathrm{bl})>0$ means the upper point merging into the bulk conduction band, then $\Domega\gtrless0$ is respectively above (bulk continuum) or below (surface state region) the merging point; $d_{3}(\bk_\mathrm{bl})<0$ similarly accounts for merging into the valence band. 
Setting $\Domega=0$, one finds $C=8t^2d_{3}(\bk_\mathrm{bl})\partial_\bk\varepsilon|_\mathrm{bl}\cdot\Dbk$. The factor $\partial_\bk\varepsilon|_\mathrm{bl}$ fixes the local chirality, i.e., the sign of the slope, of surface dispersion at the merging point: if it is increasing, say, with respect to $k_x$, and enters the conduction band ($d_{3}(\bk_\mathrm{bl})>0$), its left ($\Delta k_x<0$) must be in the bulk continuum with $C<0$ while the right ($\Delta k_x>0$) is outside. These are all observed in, e.g., Fig.~\ref{Fig:Aomegakxkz_main}(A) and also Fig.~\ref{Fig:Aomegakxkz_typeII_SM}(A) for the type-II case. Therefore, compared to $\Delta\omega$ deviation involving $\sgn(d_{3}(\bk_\mathrm{bl}))$, one has an extra sign of $\partial_\bk\varepsilon|_\mathrm{bl}$ along some $\Dbk$ direction; together they \textit{fully} determine the local merging configuration.

In the very vicinity of the WPs, as shown in \eqref{eq:A03asymptote_wp_main}, now any $\Domega$ causes $C<0$, reflecting that pure energy deviation from a WP always enters the bulk Weyl cone however tilted, as shown in, e.g., the type-II Fig.~\ref{Fig:Aomegakxkz_typeII_SM}(B). 
When $\Domega=0$, $\Lambda$ determines the sign of $C$ and is positive-definite unless the velocity tilting $\partial_\bk d_0\vert_\mathrm{wp}$ is large enough. This directly corresponds to the distinction between type-I and type-II WSMs as shown in Fig.~\ref{Fig:Adkxdkzbl_main}(B), since $\Lambda$ with a negative eigenvalue means $C<0$ along a certain range of $\Dbk$ direction, i.e., such momentum deviation enters immediately the bulk electron/hole pockets once it leaves the WP. For the present model with $|t_2|<|t|$, the two eigenvalues of the  principal axes of $\Lambda$ are $\lambda_1=4t^2t_1^2,\lambda_2=4t^2\sin^2{\bark_z}(t^2-t_2^2-D_z^2)$ respectively along $k_x,k_z$. Hence, large $D_z$ and deviation $\Delta k_z$ can enter the bulk continuum while $\Delta k_x$ does not in the spectroscopy.

\begin{acknowledgments}
This work was supported by JSPS KAKENHI (No.~18H03676) and JST CREST (No.~JPMJCR1874). X.-X.Z. was partially supported by RIKEN Special Postdoctoral Researcher Program.
\end{acknowledgments}\mycomment{\Yinyang}

%\nocite{*}

\bibliography{reference.bib}  % The references (bibliography) information are stored in the file named "Bibliography.bib"
\let\addcontentsline\oldaddcontentsline% Restore \addcontentsline

\newpage
\onecolumngrid
\newpage
{
	\center \bf \large 
	Supplemental Information \\
	\large for ``\newtitle"\vspace*{0.1cm}\\ 
	\vspace*{0.5cm}
	%\newauthor
}
\begin{center}
    %\getauthor \\
	Xiao-Xiao Zhang$^{1,2}$ and Naoto Nagaosa$^2$\\
	\vspace*{0.15cm}
    \small{$^1$\textit{Wuhan National High Magnetic Field Center and School of Physics, Huazhong University of Science and Technology, Wuhan 430074, China}}\\
	\small{$^2$\textit{RIKEN Center for Emergent Matter Science (CEMS), Wako, Saitama 351-0198, Japan}}\\
	\vspace*{0.25cm}	
\end{center}

%\twocolumngrid	

\tableofcontents

% %\clearpage
% %\appendix
% \setcounter{equation}{0}
% \setcounter{figure}{0}
% \setcounter{table}{0}
% \setcounter{page}{1}
% %\renewcommand{\theequation}{S\arabic{equation}}
% \renewcommand{\thefigure}{S\arabic{figure}}
% \renewcommand{\bibnumfmt}[1]{[S#1]}
% %\renewcommand{\citenumfont}[1]{S#1}

%%%%%%%%%% Merge with supplemental materials %%%%%%%%%%
%%%%%%%%% Prefix a "S" to all equations, figures, tables and reset the counter %%%%%%%%%%
%\appendix
\setcounter{equation}{0}
\setcounter{figure}{0}
\setcounter{table}{0}
\setcounter{page}{1}
%\makeatletter
\renewcommand{\theequation}{S\arabic{equation}}
\renewcommand{\thefigure}{S\arabic{figure}}
\renewcommand{\theHtable}{Supplement.\thetable}
\renewcommand{\theHfigure}{Supplement.\thefigure}
\renewcommand{\bibnumfmt}[1]{[S#1]}
\renewcommand{\citenumfont}[1]{S#1}
%%%%%%%%% Prefix a "S" to all equations, figures, tables and reset the counter %%%%%%%%%%

\section{Formulation of general model Hamiltonians}\label{SM:Hamiltonian}
One can consider more general WSM Hamiltonians such as
\begin{equation}\label{eq:H}
\begin{split}
    \cH(\bk)&= \sum_{i=x,y,z} [D_i (1-\cos{k_i})\sigma_0 +t_i(\cos{k_i}-\cos{\bark_i})\sigma_1]\\
    &+(v_0+v_x\sin{k_x}+v_y\sin{k_y}+v_z\cos{k_z})\sigma_2\\
    &+ (u_0+u_x\sin{k_x}+u_z\cos{k_z})\sigma_3,
\end{split}
\end{equation}
which also can bear two WPs at $\bbark^\pm=(\bark_x,\bark_y,\pm \bark_z)$ under the following condition: $|u_{0,z}|<|u_x|,|v_{0,x,z}|<|v_y|$ and $\bark_x=\sin^{-1}\frac{u_0+u_z\cos{\bark_z}}{-u_x},\bark_y=\sin^{-1}\frac{v_0+v_x\sin{\bark_x}+v_z\cos{\bark_z}}{-v_y}$. 
Following the same procedure outlined in \textit{Methods}, %to make the recursive Eq.~\eqref{eq:H_recursive}, 
one obtains $h=d_\alpha(\bk)\sigma_\alpha$
with 
\begin{equation}
\begin{split}
d_0=\sum_{i=x,z} D_i (1-\cos{k_i}),&\quad
d_{1}=\sum_{i=x,z}t_i\cos{k_i}-\sum_{i=x,y,z}t_i\cos{\bark_i},\\
d_{2}=v_0+v_x\sin{k_x}+v_z\cos{k_z},&\quad
d_3=u_0+u_x\sin{k_x}+u_z\cos{k_z}
\end{split}
\end{equation}
and $V=\frac{1}{2}(-D_y\sigma_0+\ii v_y\sigma_2+t_y\sigma_1)$,
which is not immediately analytically tractable. 
We firstly notice that $D_y$, which introduces extra spin-independent dispersion along $k_y$, does not fundamentally affect the discussion of the sample with a surface normal to $\hat{y}$-axis and can be safely neglected; note that more important $D_x,D_z$ for the surface in $xz$-plane, which lead to the distortion of the surface state and especially the Fermi arc curvature, are included. We secondly put $t_y=v_y$ without much loss of generality, which insignificantly affects the Fermi velocity along $k_y$ and now serves as the coupling strength between the surface layer and the bulk. With these assumptions, the analytical solution can be obtained in essentially a similar way. In fact, such a type of solution mainly relies on making the hoppings in the normal direction simple, which is $\hat{y}$ in our setting. On the other hand, there is basically no constraint for the in-plane directions: $d_\alpha$ can be any function of $k_x,k_z$ and for instance, it is not limited to nearest neighbor hoppings as in Eq.~\eqref{eq:H}.
Therefore, a large class of WSM Hamiltonians can be exactly treated similarly and on this more general basis, one can physically expect that those beyond such an analytical form also share similar properties. %results as those presented in the main text.

\section{Noncentrosymmetric WSM}\label{SM:Pbreak}

\begin{figure*}[!htbp]
\centering
\includegraphics[width=17.8cm]{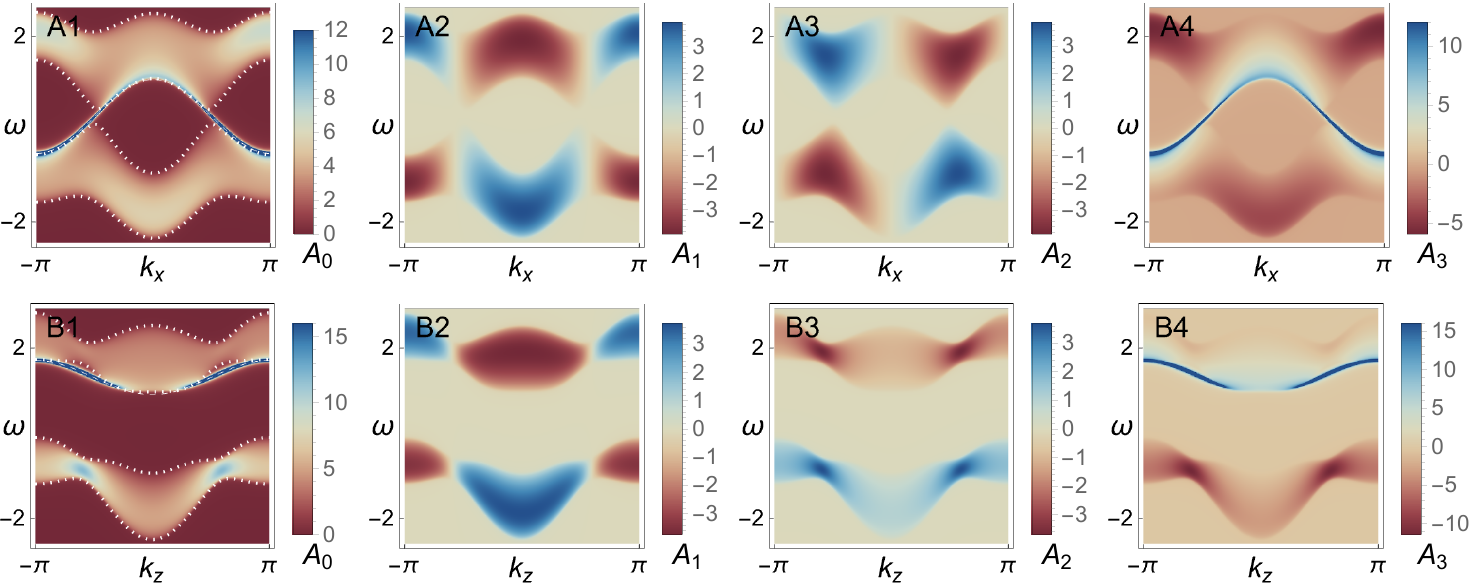}
\caption{Spin-resolved ARPES signals of the surface of a noncentrosymmetric WSM in (A) the $(\omega,k_x)$-plane at $k_z=0.2\pi$ and (B) the $(\omega,k_z)$-plane at $k_x=0.15\pi$. Spectral function $A_{0,1,2,3}$ successively in the charge and three spin channels for the WSM with parameters $k_w=\pi/2,t=t_1=1,t_2=0.8,D_x=0.2,D_z=0.4$. 
The deep blue surface state in panels (A1,A4) and (B1,B4) respectively corresponds to the green and red lines in Fig.~\ref{fig:SSreg-band_Pbreak_SM}(B). The intermittence of surface state can be clearly seen around the $\Gamma$-point in panels (A1,A4) and (B1,B4), manifesting the multiply connected shape of the surface state constraint region in Fig.~\ref{fig:SSreg-band_Pbreak_SM}(A).
Exemplified in the $A_0$ channel, it tangentially merges into the projected bulk states bounded by the dotted lines; for clarity, the full surface state dispersion beyond the momentum constraint region is indicated by dashed lines. 
}\label{Fig:Aomegakxkz_Pbreak_SM}
\end{figure*}

\begin{figure}[!tbhp]
\centering
\includegraphics[width=10.7cm]{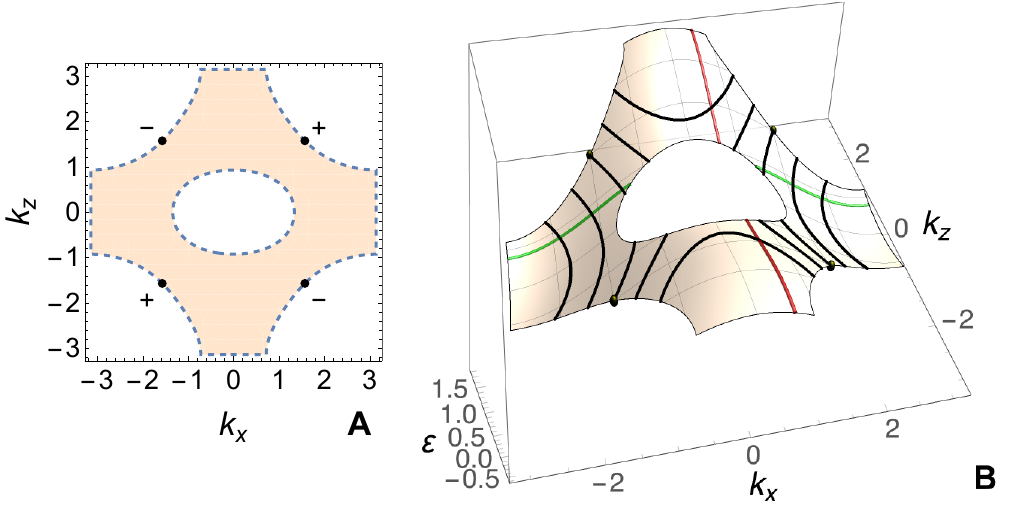}
\caption{Topological surface state on the top surface of a noncentrosymmetric WSM with two pairs of WPs. (A) The orange multiply connected momentum constraint region of the surface state bears a hole surrounding the $\Gamma$-point in the 2D surface Brillouin zone (BZ). WP charges $\pm$ are noted. (B) The energy dispersion $\varepsilon(\bk)$ of the surface state restricted in the constraint region. The projected pairs of WPs are indicated by black dots. Green and red lines are momentum cuts of the surface state at fixed $k_z$ or $k_x$, respectively corresponding to Fig.~\ref{Fig:Aomegakxkz_Pbreak_SM}(A,B).
Black lines exemplify the Fermi arcs at various Fermi energies from low to high, successively corresponding to panels in Fig.~\ref{Fig:AkxkzArc_Pbreak_SM}. Parameters same as Fig.~\ref{Fig:Aomegakxkz_Pbreak_SM}.
}
\label{fig:SSreg-band_Pbreak_SM}
\end{figure}

\begin{figure*}[!htbp]
\centering
\includegraphics[width=17.8cm]{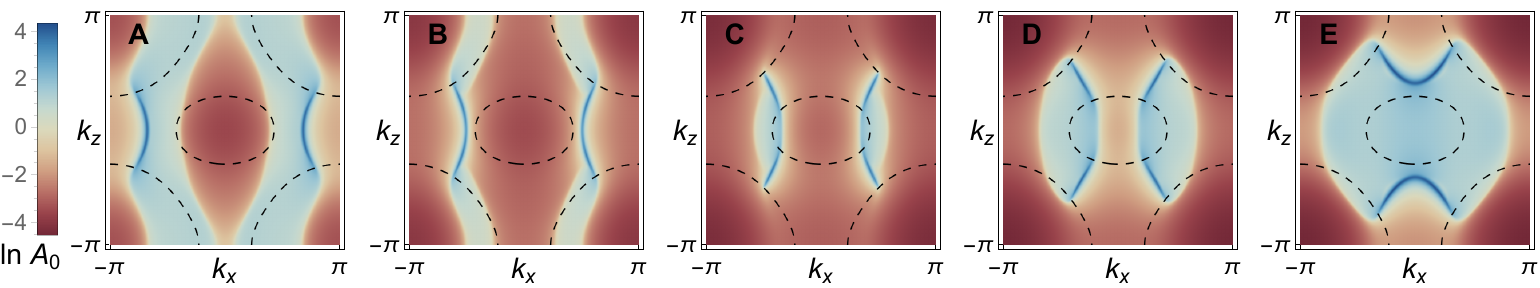}
\caption{ARPES signal log-scaled in the charge channel $A_{0}$ of the surface of a noncentrosymmetric WSM at several isoenergy planes showing the Fermi arc surface state (deep or dark blue) merging into the bulk. Panels (A-E), energies from low to high, correspond to the Fermi arcs as black lines in Fig.~\ref{fig:SSreg-band_Pbreak_SM}(B); panel (C) is the case when $E_F$ is pinned to the WP. Note that panels (C,D) show Fermi arcs doubled in number than WP pairs; panel (E) shows that the alignment of Fermi arcs is finally changed to be along $k_x$-direction, compared to panels (A,B). 
Dashed lines indicate the boundary of the surface state constraint region, corresponding to Fig.~\ref{fig:SSreg-band_Pbreak_SM}(A). 
Parameters same as Fig.~\ref{Fig:Aomegakxkz_Pbreak_SM}.
}\label{Fig:AkxkzArc_Pbreak_SM}
\end{figure*}

The Hamiltonian Eq.~\eqref{eq:H_main} in the main text breaks the time-reversal symmetry $\cT$ and is mainly considered to have a real spin degree of freedom. However, at the level of general model construction, Pauli matrices can either stand for a real spin or a pseudospin and our formalism does not depend on this choice. To consider inversion symmetry $\cP$, for a real spin, the inversion operation makes $\bk\rightarrow-\bk$; for a pseudospin due to orbital degree of freedom, the inversion additionally has the unitary matrix operation, which we consider to be $M_\cP=\sigma_1$ here. In this sense, Hamiltonian Eq.~\eqref{eq:H_main} in general breaks inversion symmetry, which can be easily seen from its asymmetric WP location with respect to inversion. For the pseudospin case, special $t_2=0$ makes it inversion symmetric. In any case, $\cT$-breaking determines the minimal number of WP pair is one.

We can otherwise consider the $\cP$-broken WSM but with $\cT$ preserved, which has a minimum of two pairs of WPs. To this end, %in Eq.~\eqref{eq:h} we can, for instance, set $d_2=t_2\sin{k_z},d_3=t_1\cos{k_x}$ and . 
we can study, for instance, the following model with $\bbark=(\frac{\pi}{2},\sin^{-1}{\frac{t_2}{t}},k_w)$
\begin{equation}\label{eq:H_P1}
\begin{split}
    &\cH(\bk)= [D_x (1-\cos{k_x})+D_z (1-\cos{k_z})]\sigma_0+t_1\cos{k_x}\sigma_3 \\
    &+(t\sin{k_y}-t_2\sin{k_x})\sigma_2
     -t \sum_{i=x,y,z}(\cos{k_i}-\cos{\bark_i})\sigma_1, 
\end{split}
\end{equation}
which is also the model used in Fig.~\ref{fig:SSreg-band_main}(C1,C2). It respects $\cT$ as a pseudospin model with the $\cT$ operation given by complex conjugation. The two pairs of WPs are located at $(q\bark_x,q\bark_y,pq\bark_z)$ with $p=\pm1$ and $q=\pm1$ independently. We restrict the range of $\sin^{-1}$ to $[-\frac{\pi}{2},\frac{\pi}{2}]$ throughout the study. The $\cT$-related monopole pair ($p=1$) and antimonopole pair ($p=-1$) are the $q=\pm1$ pairs; the monopole-antimonopole pairs for $p=\pm1$ with given $q$ are connected by topological surface states with Fermi arcs. 

In the same manner as the $\cT$-broken model in the main text, we present for this WSM model the ARPES signal in Fig.~\ref{Fig:Aomegakxkz_Pbreak_SM}, the surface state information in Fig.~\ref{fig:SSreg-band_Pbreak_SM}, and the Fermi arc spectroscopy in Fig.~\ref{Fig:AkxkzArc_Pbreak_SM}. The blue or red spots in the lower branch in Fig.~\ref{Fig:Aomegakxkz_Pbreak_SM}(B1,B3,B4) also exemplify the discussion in Sec.~\ref{SM:g_singularity} as they are slightly near the divergence. The absence of such spots in Fig.~\ref{Fig:Aomegakxkz_Pbreak_SM}(B2) is because the Green's function component $b_1$ exactly vanishes there as per the Hamiltonian.

\begin{figure*}[!htbp]
\centering
\includegraphics[width=17.8cm]{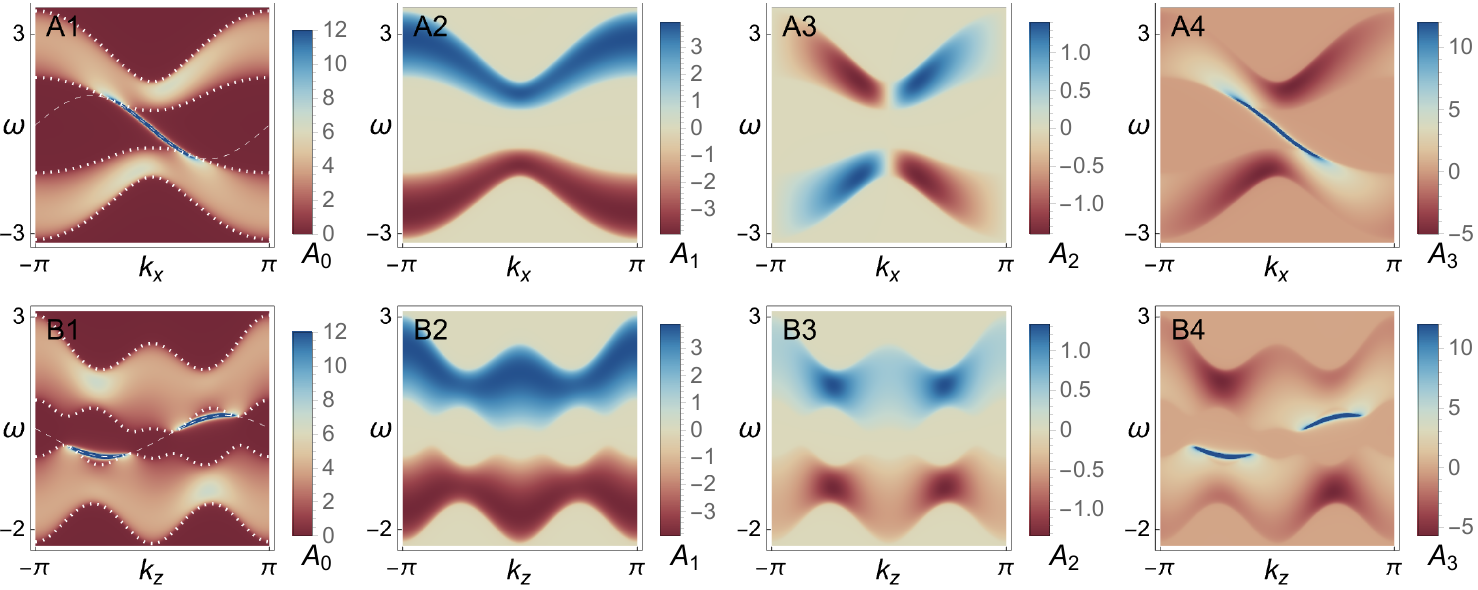}
\caption{Spin-resolved ARPES signals of the surface of a noncentrosymmetric WSM in (A) the $(\omega,k_x)$-plane at $k_z=-0.4\pi$ and (B) the $(\omega,k_z)$-plane at $k_x=0.15\pi$. Spectral function $A_{0,1,2,3}$ successively in the charge and three spin channels for the WSM with parameters $\bark_{zL}=\pi/6,\bark_{zR}=3\pi/4,t=t_1=1,t_2=0.4,t_3=0,D=0.2,D_x=0$. 
The deep blue surface state in panels (A1,A4) and (B1,B4) respectively corresponds to the green and red lines in Fig.~\ref{fig:SSreg-band_Pbreak2_SM}(B). 
Exemplified in the $A_0$ channel, it tangentially merges into the projected bulk states bounded by the dotted lines; for clarity, the full surface state dispersion beyond the momentum constraint region is indicated by dashed lines. 
}\label{Fig:Aomegakxkz_Pbreak2_SM}
\end{figure*}

\begin{figure}[!tbhp]
\centering
\includegraphics[width=10.7cm]{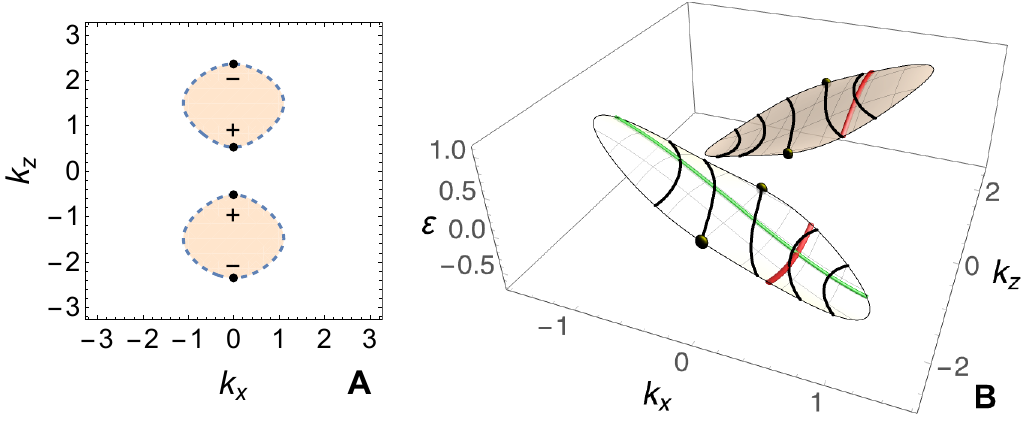}
\caption{Topological surface state on the top surface of a noncentrosymmetric WSM with two pairs of WPs along the $k_z$ direction. (A) The orange momentum constraint region of the surface state bears two disconnected parts in the 2D surface Brillouin zone. WP charges $\pm$ are noted. (B) The energy dispersion $\varepsilon(\bk)$ of the surface state restricted in the constraint region. The projected pairs of WPs are indicated by black dots. Green and red lines are momentum cuts of the surface state at fixed $k_z$ or $k_x$, respectively corresponding to Fig.~\ref{Fig:Aomegakxkz_Pbreak2_SM}(A,B).
Black lines exemplify the Fermi arcs at various Fermi energies from low to high, successively corresponding to panels in Fig.~\ref{Fig:AkxkzArc_Pbreak2_SM}. Parameters same as Fig.~\ref{Fig:Aomegakxkz_Pbreak2_SM}.
}
\label{fig:SSreg-band_Pbreak2_SM}
\end{figure}

\begin{figure*}[!htbp]
\centering
\includegraphics[width=17.8cm]{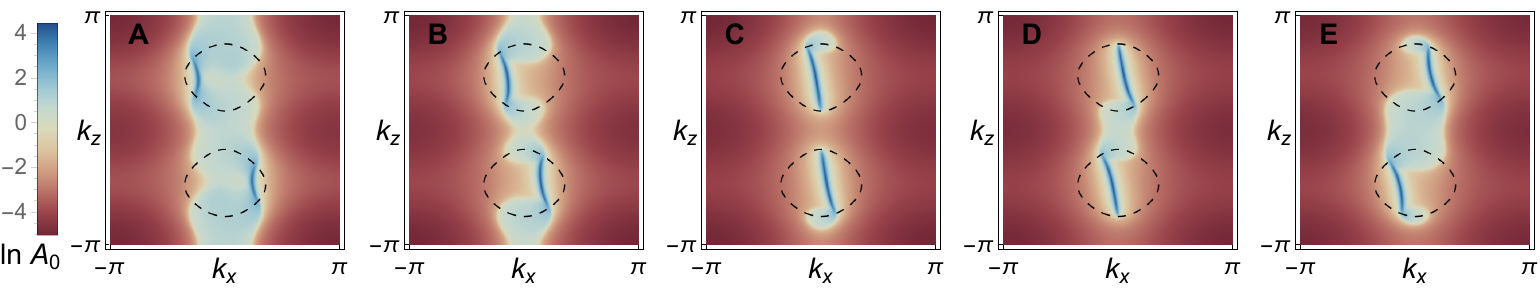}
\caption{ARPES signal log-scaled in the charge channel $A_{0}$ of the surface of a noncentrosymmetric WSM at several energy cuts showing the Fermi arc surface state (deep or dark blue) merging into the bulk. Panels (A-E), energies from low to high, correspond to the Fermi arcs as black lines in Fig.~\ref{fig:SSreg-band_Pbreak2_SM}(B); panels (C,D) are when $E_F$ is respectively pinned to the inner and outer WP pairs. 
Dashed lines indicate the boundary of the surface state constraint region, corresponding to Fig.~\ref{fig:SSreg-band_Pbreak2_SM}(A).
Parameters same as Fig.~\ref{Fig:Aomegakxkz_Pbreak2_SM}.
}\label{Fig:AkxkzArc_Pbreak2_SM}
\end{figure*}

Another $\cP$-broken WSM model is\cite{XXZ:Luttinger} 
\begin{equation}\label{eq:H_P2}
\begin{split}
    &\cH(\bk)= [D(1-\cos{k_x}\cos{k_z})+D_x\sin{2k_x}]\sigma_0+t_1\sin{k_x}\sin{k_z}\sigma_3\\
    &+ (t\sin{k_y}+t_2\sin{k_x}-t_3)\sigma_2\\
    &+
     [(\cos{k_z}-\cos{\bark_{zL}})(\cos{k_z}-\cos{\bark_{zR}})-t \sum_{i=x,y}(\cos{k_i}-\cos{\bark_i})]\sigma_1,
\end{split}
\end{equation}
given $\bark_x=0,\bark_y=\sin^{-1}\frac{t_3}{t}$. Note that for simplicity we keep using the same $\sigma_0$ part in the previous calculations, which is certainly not a requirement; instead, as exemplified in this model, it can be any other functions of $\bk$. %for the symmetry of the present case. 
This system has two pairs of WPs all aligned along the $k_z$-axis at $(\bark_x,\bark_y,\pm \bark_{zL})$ and $(\bark_x,\bark_y,\pm \bark_{zR})$, where the former forms a monopole pair (WP charge $+1$) and the latter forms an antimonopole pair (WP charge $-1$), provided that $0<\bark_{zL}<\bark_{zR}<\pi$. When $t_3=D_x=0$, it further respects $\cT$-symmetry, which is what we will focus on. Also, finite $D$ makes the inner two WPs lower in energy than the outer two WPs.

In the same manner as the $\cT$-broken model in the main text, we present for this WSM model the ARPES signal in Fig.~\ref{Fig:Aomegakxkz_Pbreak2_SM}, the surface state information in Fig.~\ref{fig:SSreg-band_Pbreak2_SM}, and the Fermi arc spectroscopy in Fig.~\ref{Fig:AkxkzArc_Pbreak2_SM}. As expected from the WP charges, we observe in Fig.~\ref{fig:SSreg-band_Pbreak2_SM} two separate constraint regions with each connecting a monopole-antimonopole pair. In this case, although the constraint region is not simply connected in the first place, there is not any inner hole inside the constraint region, we hence find the same number of Fermi arcs as that of WP pairs in Fig.~\ref{Fig:AkxkzArc_Pbreak2_SM}.

\section{Time-reversal-breaking WSM with more pairs of WPs}\label{SM:Tbreak}

\begin{figure*}[!htbp]
\centering
\includegraphics[width=17.8cm]{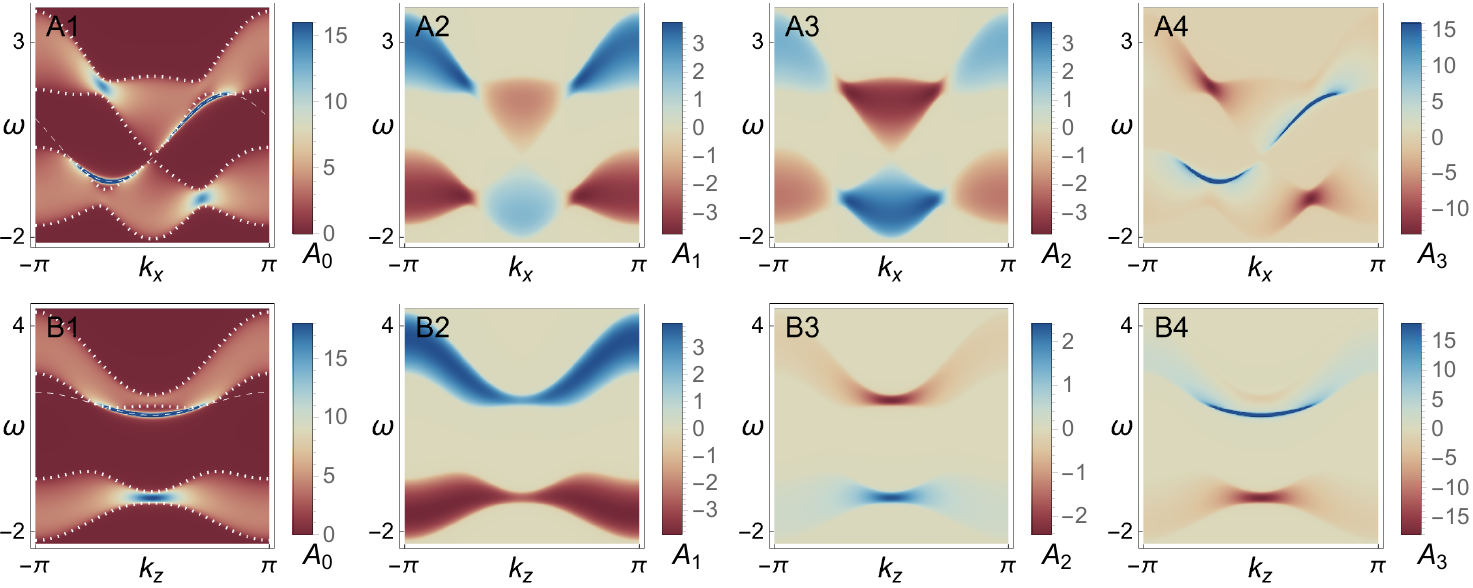}
\caption{Spin-resolved ARPES signals of the surface of a $\cT,\cP$-broken WSM in (A) the $(\omega,k_x)$-plane at $k_z=0.5$ and (B) the $(\omega,k_z)$-plane at $k_x=1.4$. Spectral function $A_{0,1,2,3}$ successively in the charge and three spin channels for the WSM with parameters $k_w=\pi/2,t=t_1=1,t_2=0.94,D_x=0.5,D_z=0.33$. 
The deep blue surface state in panels (A1,A4) and (B1,B4) respectively corresponds to the green and red lines in Fig.~\ref{fig:SSreg-band_Tbreak2_SM}(B). The intermittence of surface state can be clearly seen around the $\Gamma$-point in panels (A1,A4), manifesting the multiply connected shape of the surface state constraint region in Fig.~\ref{fig:SSreg-band_Tbreak2_SM}(A).
Exemplified in the $A_0$ channel, it tangentially merges into the projected bulk states bounded by the dotted lines; for clarity, the full surface state dispersion beyond the momentum constraint region is indicated by dashed lines. 
}\label{Fig:Aomegakxkz_Tbreak2_SM}
\end{figure*}

\begin{figure}[!tbhp]
\centering
\includegraphics[width=10.7cm]{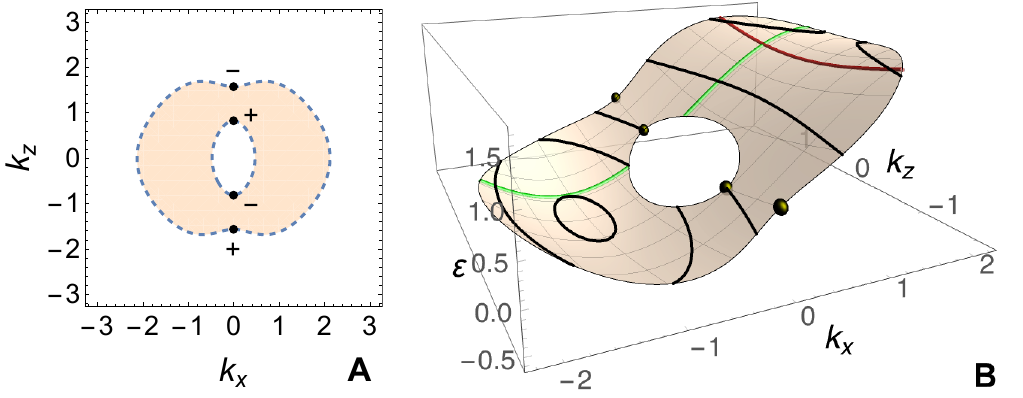}
\caption{Topological surface state on the top surface of a $\cT,\cP$-broken WSM with two pairs of WPs at different energies. (A) The orange momentum constraint region of the surface state bears an inner hole and is multiply connected. WP charges $\pm$ are noted. (B) The energy dispersion $\varepsilon(\bk)$ of the surface state restricted in the constraint region. The projected pairs of WPs are indicated by black dots. Green and red lines are momentum cuts of the surface state at fixed $k_z$ or $k_x$, respectively corresponding to Fig.~\ref{Fig:Aomegakxkz_Tbreak2_SM}(A,B).
Black lines exemplify the Fermi arcs at various Fermi energies from low to high, successively corresponding to panels in Fig.~\ref{Fig:AkxkzArc_Tbreak2_SM}. Note that the inner borderline due to the hole passes through a pair of WPs, in contrast to Fig.~\ref{fig:SSreg-band_Pbreak_SM}.
Parameters same as Fig.~\ref{Fig:Aomegakxkz_Tbreak2_SM}.
}
\label{fig:SSreg-band_Tbreak2_SM}
\end{figure}

\begin{figure*}[!htbp]
\centering
\includegraphics[width=17.8cm]{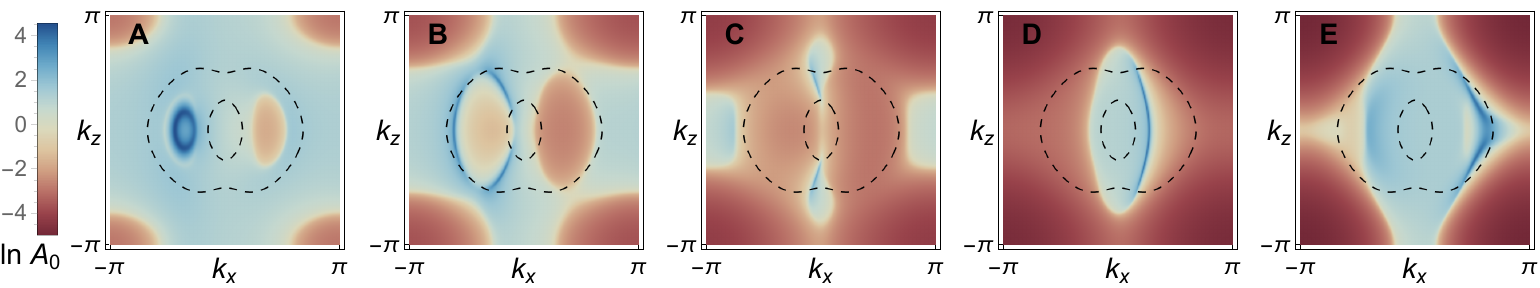}
\caption{ARPES signal log-scaled in the charge channel $A_{0}$ of the surface of a $\cT,\cP$-broken WSM at several energy cuts showing the Fermi arc surface state (deep or dark blue) merging into the bulk. Panels (A-E), energies from low to high, correspond to the Fermi arcs as black lines in Fig.~\ref{fig:SSreg-band_Tbreak2_SM}(B). Panel (C) shows Fermi arcs doubled in number than the single WP pair due to the hole; panel (B) shows the appearance of three Fermi arcs, due to the hole and also the surface state dispersion; panel (A) shows the appearance of a Fermi loop instead of Fermi arc(s). 
Dashed lines indicate the boundary of the multiply-connected surface state constraint region, corresponding to Fig.~\ref{fig:SSreg-band_Tbreak2_SM}(A). 
Parameters same as Fig.~\ref{Fig:Aomegakxkz_Tbreak2_SM}.
}\label{Fig:AkxkzArc_Tbreak2_SM}
\end{figure*}

Here we consider model Eq.~\eqref{eq:H_main} in the main text for a different set of parameters. 
In fact, it can show two WP pairs more than the minimal one and the two pairs are at different energies. This happens when $|\bark_z'|<1$ with $\bark_z'=2\cos{\bark_y}+\cos{\bark_z}$. Under this condition, the first WP pair is at $(\bark_x,\bark_y,\pm\bark_z)$ and the second pair is at $(\bark_x,\pi-\bark_y,\pm\cos^{-1}\bark_z')$. The first pair is the same as that in the main text; the second pair is created across a topological transition at $|\bark_z'|=1$. In the same manner as the figures in the main text, we present for this WSM model the ARPES signal in Fig.~\ref{Fig:Aomegakxkz_Tbreak2_SM}, the surface state information in Fig.~\ref{fig:SSreg-band_Tbreak2_SM}, and the Fermi arc spectroscopy in Fig.~\ref{Fig:AkxkzArc_Tbreak2_SM}.

%\subsection{More general Fermi arc situation}
Interestingly, the aforementioned topological transition is accompanied by the opening of an inner hole in the constraint region, as shown in Fig.~\ref{fig:SSreg-band_Tbreak2_SM}. Therefore, the present case, different from Fig.~\ref{fig:SSreg-band_main}(A,B) in the main text or Fig.~\ref{fig:SSreg-band_Pbreak_SM}, shows that the inner hole borderline can also pass through WPs. As discussed in the main text, such an inner hole increases the number of Fermi arcs as shown in Fig.~\ref{Fig:AkxkzArc_Tbreak2_SM}(C), although only one WP pair is at play at this particular energy.

\section{Configuration of Fermi arcs and surface connectivity}\label{SM:connectivity}

Here, we discuss the projected connection pattern of the WSM surface states, i.e., the configuration of Fermi arcs, especially for the case with multiple pairs of WPs. 
Firstly, it is important to note that such surface connectivity is not universally determined by the bulk band and can be affected in complex ways by various factors related to the surface, including but not limited to surface relaxation and surface reconstruction\cite{Lv2021}. 
Secondly, there still exist rich but generic features that can be captured within the present formalism. Based on the three representative models we have introduced so far and discussed below, we can accurately observe generic properties of the surface projected arc connection and, importantly, understand how the usual association of Fermi arcs
and WP pairs is unjustified.

There are mainly three factors involved: i) surface state constraint region in the momentum space, ii) surface state energy surface profile, and iii) position of the Fermi energy $E_F$ or chemical potential $\mu$. The connection pattern of the projected surface state can change when the parameters and hence these three factors of a WSM are varied. Even for a given WSM system, i.e., the former two are fixed, tuning $\mu$ will have a nontrivial influence on the Fermi arc geometry, which is, however, often overlooked in earlier studies\cite{Kim_2016}.

Fig.~\ref{fig:SSreg-band_Pbreak_SM} is based on the noncentrosymmetric WSM system Eq.~\eqref{eq:H_P1} with two pairs of WPs located at $(q\bark_x,q\bark_y,pq\bark_z)$ with $p=\pm1$ and $q=\pm1$ independently. The $\cT$-related monopole pair ($p=1$) and antimonopole pair ($p=-1$) are the $q=\pm1$ pairs; the monopole-antimonopole pairs correspond to $p=\pm1$ with given $q$.
i) As discussed in the main text, due to the multiply connected constraint region with a hole, a surface projected WP pair is \textit{not} connected by a single Fermi arc at the WP energy in Fig.~\ref{fig:SSreg-band_Pbreak_SM}(B), it instead is cut into two separate and disconnected arcs.
ii) Were it not for the hole, one Fermi arc would connect a pair of opposite WPs. However, the connection pattern, i.e., which pair is connected, is strongly affected by the shape of the energy surface and $\mu$. As shown in Fig.~\ref{fig:SSreg-band_Pbreak_SM}(B), below (above) the WP energy, the Fermi arcs follow the trend of connecting the WP pairs along $k_z$ ($k_x$). Thus, tuning the position of $\mu$ can switch between complementary connection patterns.

Fig.~\ref{fig:SSreg-band_Pbreak2_SM} is based on the noncentrosymmetric WSM system Eq.~\eqref{eq:H_P2} with two pairs of WPs all aligned along the $k_z$-axis at $(\bark_x,\bark_y,\pm \bark_{zL})$ and $(\bark_x,\bark_y,\pm \bark_{zR})$, where the former has WP charge $+1$ and the latter has WP charge $-1$. In this case with WP energies not all the same, Fermi arcs do not directly connect the projected WPs, although they follow the trend of connecting opposite WPs within each disconnected constraint region. This is thus a simpler case compared to Fig.~\ref{fig:SSreg-band_Pbreak_SM}.

Fig.~\ref{fig:SSreg-band_Tbreak2_SM} is based on the WSM system Eq.~\eqref{eq:H_main} when it exhibits two WP pairs at different energies as mentioned in Sec.~\ref{SM:Tbreak}.
This case exemplifies more unconventional properties of Fermi arc geometry. 
i) The third lowest Fermi energy cut in Fig.~\ref{fig:SSreg-band_Tbreak2_SM}(B) and the corresponding Fig.~\ref{Fig:AkxkzArc_Tbreak2_SM}(C) shows disconnected two Fermi arcs at the same energy of a WP pair. Each arc starts from one WP but merely ends at the constraint region boundary, which is even more distinct from a hole obstructing a complete arc as in Fig.~\ref{fig:SSreg-band_Pbreak_SM}.
ii) The second lowest Fermi energy cut in Fig.~\ref{fig:SSreg-band_Tbreak2_SM}(B) and the corresponding  Fig.~\ref{Fig:AkxkzArc_Tbreak2_SM}(B) further exhibit \textit{three} Fermi arcs instead of only one, which is caused by two factors. On one hand, the inner hole of the constraint region cuts the right part of the arc; on the other hand, due to the locally convex shape of the surface state band, another segment of Fermi arc also appears as the left part. iii) Further lowering $E_F$ or $\mu$, we see the appearance of a Fermi loop rather than any open Fermi arc in Fig.~\ref{Fig:AkxkzArc_Tbreak2_SM}(A) although it is entirely from the projected surface state, which is still due to the locally convex surface state band. In addition, we notice that in Fig.~\ref{Fig:AkxkzArc_Tbreak2_SM}(E), also due to the local dispersion of the surface state band and the shape of the constraint region, one Fermi arc can split into two segments, not because of a hole region. These plenty of examples show again that the appearance of the Fermi arc is a phenomenon dependent on complex factors and is not in any one-to-one correspondence with the number of WP pairs.

\section{Singularity of the surface effective Green's function due to bulk band edge touching}\label{SM:g_singularity}
From the formula Eq.~\eqref{eq:ginv&g} of surface effective Green's function $g$ in the main text, %\eqref{eq:ginv&g}, 
we note that $g$ can have a singularity when 
\begin{equation}
    d_1=d_2=0,
\end{equation}
which is not always possible and not related to the surface state resonance. The physical meaning of such a singularity can be seen from the edge formula Eq.~\eqref{eq:cE_t} of projected bulk continuum $\cE_s$ in the main text, %\eqref{eq:cE_t}, 
which now becomes 
\begin{equation}
    \cE_s=\sqrt{d_3^2+t^2}
\end{equation}
independent of $s=\pm$. For either the conduction or valence bulk band, two branches of $s$ specify the band edges from above and below. Hence, the singularity is where the degeneracy of such band edges occurs and the surface projected bulk band edge is not smooth due to such touching. Apart from such accidental singularity, we note that 
\begin{equation}
    \Det g=-\frac{K+2t^2}{2d_{12}^2t^2}
\end{equation}
is \textit{not} singular anywhere. This apparently implies no exactly on-resonance state. However, interestingly, a pole can still appear in some components of $g$, exemplified by the surface state resonance as seen from the spectral function Eq.~\eqref{eq:specA} in the charge and spin-$S_z$ channel in the main text.%\eqref{eq:specA}.

\section{Type-II WSM}\label{SM:typeII}
As mentioned in the main text, increasing the tilting effect due to $D_z$ in the Hamiltonian Eq.~\eqref{eq:H_main}, which enters the surface Hamiltonian $h(\bk)$ 
%Eq.~\eqref{eq:h} 
via $d_0(\bk)$, will eventually lead to the Lifshitz transition and reach a type-II WSM. 
In fact, given a general bulk Hamiltonian $\cH=a_\alpha(\bk)\sigma_\alpha$, one can consider the following rank-2 tensor 
\begin{equation}
    M= \partial_\bk\ba(\partial_\bk\ba)^T-\partial_\bk a_0\partial_\bk a_0 \vert_\mathrm{wp}.
\end{equation}
When $a_0$, which is identical to $d_0$ in our case, is large enough, $M$ becomes no longer positive-definite, signifying the Lifshitz transition. This intuitively means that along some certain direction $\tilde\bk$ the velocity $\bv=\frac{\partial E}{\partial \tilde\bk}\vert_\mathrm{wp}$ evaluated at the WPs changes its sign, where the band energy of the system reads $E(\bk)=a_0\pm|\ba|$.
In the main text discussion on the merging behavior, we have related this to the positivity of another tensor $\Lambda$ defined for the surface in Eq.~\eqref{eq:A03asymptote_wp_main}.
Since this is by definition a bulk condition, there is some difference worth mentioning. $\Lambda$, in a way, condenses the bulk dispersion along $k_y$-axis and misses the accurate $M(k_y)$ dependence. Hence, there exists the following rare possibility: the negative eigenvalue of $\Lambda$ (i.e., crossing some bulk pockets once the momentum leaves the projected WP in $k_x$-$k_z$ plane) is not caused by the Lifshitz transition at the WP; instead, it comes from that bulk $k_y$-dispersion crosses the WP energy at some momentum far away from the WP. Such a situation either comes from a special $a_0(\bk)$ dispersion or is that the extra crossing is at other higher-order band touchings. However, we can actually disregard this complexity because all two-band models under our consideration do not fall into such a rare situation.

Taking Eq.~\eqref{eq:H_main} as an example, we find 
\begin{equation}\label{eq:M1}
    M=
    \begin{bmatrix}
t_1^2 &  & \\
 & t^2 & tt_2\sin{\bark_z}\\
 & tt_2\sin{\bark_z} & (t^2-D_z^2)\sin^2{\bark_z}
\end{bmatrix},
\end{equation}
whose three eigenvalues are
\begin{equation}
\begin{split}
    \lambda_0=t_1^2,\quad \lambda_\pm=I \pm \sqrt{J}
\end{split}
\end{equation}
with $x=\sin{\bark_z},I=t^2(1+x^2)-D_z^2x^2,J=[(D_z^2-t^2)x^2+t^2]^2+4t^2t_2^2x^2$. One can readily prove that $\lambda_\pm>0$, i.e., a type-I case, requires 
\begin{equation}
    I^2-J=4t^2x^2(t^2-t_2^2-D_z^2)>0,
\end{equation} 
which is identical to the condition mentioned in the main text.
In addition, Eq.~\eqref{eq:M1} indicates that the bulk Fermi pockets connected to WPs are within a range of direction in between $k_y,k_z$. This is why even $D_z=t$ suffices to make the system type-II, due to the presence of $t_2$ related to $k_y$ dispersion.

\begin{figure*}[!htbp]
\centering
\includegraphics[width=17.8cm]{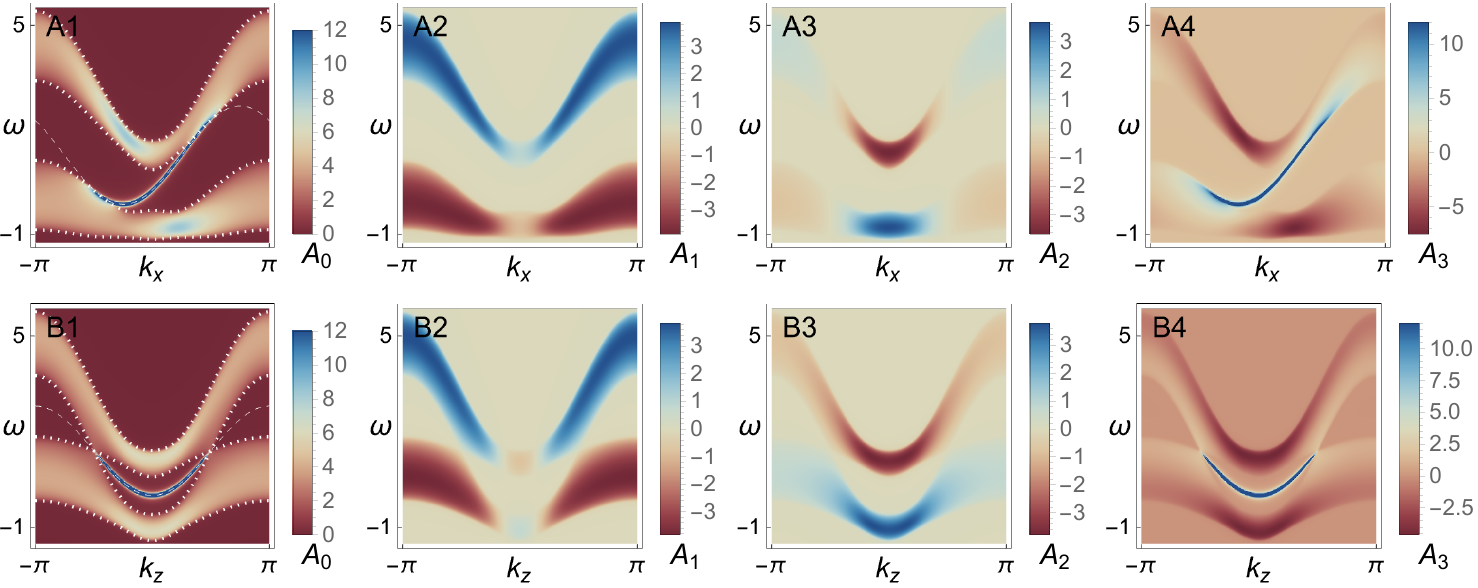}
\caption{Spin-resolved ARPES signals of the surface of a type-II WSM in (A) the $(\omega,k_x)$-plane at $k_z=\pi/5$ and (B) the $(\omega,k_z)$-plane at $k_x=0$. Spectral function $A_{0,1,2,3}$ successively in the charge and three spin channels for the WSM with parameters $k_w=\pi/2,t=t_1=1,t_2=0.4,D_x=1,D_z=1.4$. 
A deep blue surface state (resonance intensity clipped to enhance the overall visibility) is seen only and identically in the $A_{0,3}$ channels, signifying the charge-spin locking between them. 
Such a surface state in panels (A1,A4) and (B1,B4) respectively corresponds to the green and red lines in Fig.~\ref{fig:SSreg-band_typeII_SM}(B).
Exemplified in the $A_0$ channel, it tangentially merges into the projected bulk states bounded by the dotted lines; for clarity, the full surface state dispersion beyond the momentum constraint region is indicated by dashed lines. A pair of type-II WP touchings can be seen in (B1,B4), where the surface state ends.
}\label{Fig:Aomegakxkz_typeII_SM}
\end{figure*}

\begin{figure}[!tbhp]
\centering
\includegraphics[width=10.7cm]{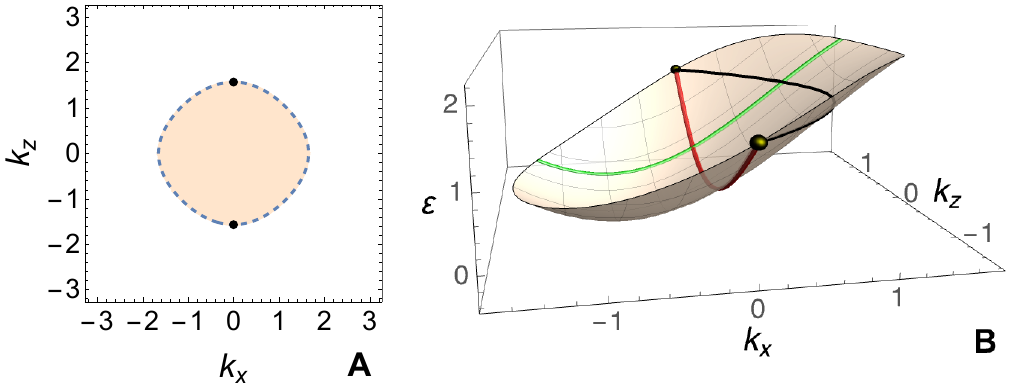}
\caption{Topological surface state on the top surface of a type-II WSM. (A) The orange momentum constraint region of the surface state in the 2D surface Brillouin zone (BZ). (B) The energy dispersion $\varepsilon(\bk)$ of the surface state restricted in the constraint region. The projected pair of WPs is indicated by black dots. Green and red lines are momentum cuts of the surface state at fixed $k_z$ or $k_x$, respectively corresponding to Fig.~\ref{Fig:Aomegakxkz_typeII_SM}(A,B).
Black line exemplifies the Fermi arc that connects the WPs. Parameters same as Fig.~\ref{Fig:Aomegakxkz_typeII_SM}.
}
\label{fig:SSreg-band_typeII_SM}
\end{figure}

It is also interesting to note that such type-II tilting only affects the surface state dispersion $\varepsilon(\bk)$ in Eq.~\eqref{eq:Esurface}, which depends on $d_0$. It neither alters the surface state constraint region Eq.~\eqref{eq:constraint} nor the intensity profile $R(\bk)$ of Eq.~\eqref{eq:A03ss_in} of the surface state resonance, both of which depend on $d_1$ and $d_2$ as given in the main text. As shown in Fig.~\ref{fig:SSreg-band_typeII_SM}(A), the surface state constraint region is the same as Fig.~\ref{fig:SSreg-band_main}(B1) in the main text. We also show its typical spectral function plot in Fig.~\ref{Fig:Aomegakxkz_typeII_SM} in the same manner as Fig.~\ref{Fig:Aomegakxkz_main} in the main text. Here, in Fig.~\ref{Fig:Aomegakxkz_typeII_SM}(B), we set the momentum cut at $k_x=\bark_x$; therefore, one can clearly see the type-II WPs with the characteristic velocity tilting, where the surface state also ends.

\end{document}